\documentclass[aps,prd,10pt,notitlepage,nofootinbib,superscriptaddress]{revtex4-1}

\usepackage[utf8]{inputenc}
\usepackage{amsmath}
\usepackage{amssymb}
\usepackage{amsfonts}
\usepackage{newtxtext,newtxmath}
\usepackage{bm}
\usepackage{graphicx}
\usepackage[usenames,dvipsnames]{xcolor}
\usepackage{color}
\usepackage[colorlinks=true,linkcolor=blue,urlcolor=blue,citecolor=blue]{hyperref}

\usepackage{slashed}
\usepackage[english]{babel}
\usepackage{dcolumn}
\usepackage{blindtext}
\usepackage{epsfig}
\usepackage{pifont}
\usepackage{dsfont,mathrsfs}
\usepackage{cancel}
\usepackage{bigints}
\usepackage{accents}
\usepackage{soul}
\usepackage{multirow}
\usepackage{simpler-wick}
\usepackage{enumerate}

\newcommand{\nn}{\nonumber}

\newcommand{\ensembleaverage}[1]{\left\langle#1\right\rangle}

\newcommand{\MB}[1]{\left|#1\right|}
\newcommand{\FB}[1]{\left(#1\right)}
\newcommand{\fb}[1]{(#1)}
\newcommand{\SB}[1]{\left\{#1\right\}}
\newcommand{\TB}[1]{\left[#1\right]}

\newcommand{\mcT}{\mathcal{T}}
\newcommand{\mcTc}{\mathcal{T}_C}
\newcommand{\mcN}{\mathcal{N}}

\newcommand{\scrD}{\mathscr{D}}

\newcommand{\munu}{{\mu\nu}}
\newcommand{\numu}{{\nu\mu}}

\newcommand{\IM}{\text{Im}}

\newcommand{\Tr}{\text{Tr}}

\newcommand{\pll}{\parallel}

\newcommand{\kpll}{k_\parallel}

\newcommand{\ppll}{p_\parallel}

\newcommand{\kper}{k_\perp}

\newcommand{\gpll}{g_\parallel}
\newcommand{\gper}{g_\perp}

\newcommand{\del}{\partial}

\newcommand{\psibar}{\overline{\psi}}

\newcommand{\Dtil}{\tilde{D}}

\newcommand{\be}{\begin{equation}}
\newcommand{\ee}{\end{equation}}
\newcommand{\bea}{\begin{eqnarray}}
\newcommand{\eea}{\end{eqnarray}}
\newcommand{\ba}[1]{\begin{array}{#1}}
\newcommand{\ea}{\end{array}}

\newcommand{\om}{\omega}

\newcommand{\vk}{\vec k}

\begin{document}

\title{Kubo estimation of the electrical conductivity for a hot relativistic fluid in the presence of a magnetic field}

\author{Sarthak Satapathy}
\email{sarthaks@iitbhilai.ac.in}
\affiliation{Indian Institute of Technology Bhilai, GEC Campus, Sejbahar, Raipur - 492015, Chhattisgarh, India}

\author{Snigdha Ghosh}
\email{snigdha.physics@gmail.com}
\thanks{Corresponding Author}
\affiliation{Government General Degree College Kharagpur-II, Paschim Medinipur - 721149, West Bengal, India}

\author{Sabyasachi Ghosh}
\email{sabyaphy@gmail.com}
\affiliation{Indian Institute of Technology Bhilai, GEC Campus, Sejbahar, Raipur - 492015, Chhattisgarh, India}

\begin{abstract}
We have explored the multi-component structure of electrical conductivity of relativistic Fermionic and Bosonic fluid in presence of magnetic field by using Kubo approach. This is done by explicitly evaluating the thermo-magnetic vector current spectral functions using the real time formalism of finite temperature field theory and the Schwinger proper time formalism. In absence of magnetic field, the one-loop diagramatic representation of Kubo expression of any transport coefficients is exactly same with relaxation time approximation (RTA) based expression, but this equality does not hold for finite magnetic field picture due to lacking of proper implementation of quantum effect in latter approach. We have shown this discrepancy for particular transport coefficient - electrical conductivity, whose starting point in Kubo approach will be electromagnetic current-current correlator and its one-loop skeleton diagram carrying two scalar/Dirac propagators for scalar/Dirac fluid. Through a numerical comparison between RTA and Kubo expressions of conductivity components (parallel and perpendicular), we have attempted to interpret detail quantum field theoretical effect, contained by Kubo expression but not by RTA expression. In classical RTA expression we get magnetic field independent parallel conductivity due to zero Lorentz force but in field theoretical Kubo expression, it decreases and increases with the magnetic field for scalar and Dirac medium respectively due to Landau quantization effect. This parallel component of conductivity can be interpreted as zero momentum limit of quantum fluctuation with same Landau level internal lines. While for perpendicular component of conductivity, fluctuation with Landau level differences $\pm 1$ are noticed, which might be a new realization of transportation in field theoretical sector.    
\end{abstract}

\maketitle

\section{INTRODUCTION}
Research on Quark Gluon Plasma (QGP), which can be produced in heavy ion collision (HIC) experiment facilities is appeared to be a mature branch of high energy physics, where a broad band of basic physics from classical mechanics to quantum mechanics to statistical mechanics to quantum field theory are largely cultivated. As interesting as it is, the creation of very strong magnetic fields of the order $10^{19}-10^{20}$ G during non-central or asymmetric HICs has thrown a plethora of questions requiring a careful study. Novel phenomena such as chiral magnetic effect~\cite{Fukushima:2008xe}, magnetic catalysis~\cite{Shovkovy:2012zn} and inverse magnetic catalysis~\cite{Bali:2012zg,Bali:2014kia,Mueller:2014tea} ask for a deep understanding of the underlying theoretical aspects of QGP. An extensive discussion of the effects of magnetic field on hot Quantum Chromodynamics (QCD) matter has been discussed in Ref~\cite{Kharzeev:2012ph}. In nature, strong magnetic fields of the order $10^{12}-10^{13}$ G in neutron stars and an even higher magnitude of $10^{15}-10^{16}$ G in magnetars are known to exist~\cite{Konar:2017kty,Potekhin:1999ur}. These values are smaller as compared to values of the magnetic fields created at LHC and RHIC. Therefore HIC provides us with an unique opportunity to investigate properties of hot QCD matter under the influence of magnetic fields.

Being an important quantity, transport coefficients of QGP is also suspected to be modified due to strong magnetic field. Their microscopic estimations become quite important as they enter as input of evolving QGP. Based on that interest, microscopic calculations of transport coefficients like shear viscosity~\cite{Li:2017tgi,Tuchin:2011jw,Ghosh:2018cxb,Dey:2019axu,Dey:2019vkn,Dash:2020vxk,Chen:2019usj,Denicol:2018rbw,Mohanty:2018eja}, bulk viscosity~\cite{Hattori:2017qih,Huang:2009ue,Agasian:2013wta,Agasian:2011st,Kurian:2018dbn} and electrical conductivity~\cite{Dey:2019axu,Hattori:2016lqx,Hattori:2016cnt,Nam:2012sg,Hess2015,Das:2019wjg,Das:2019ppb,Kalikotay:2020snc,Kerbikov:2014ofa,Kurian:2018qwb,Kurian:2017yxj,Feng:2017tsh,Fukushima:2017lvb,Harutyunyan:2016rxm} for the relativistic systems in presence of magnetic field are rigorously studied in recent time. Relaxation Time Approximation (RTA) of kinetic theory and Kubo formalism are the two frequently used approaches for microscopic calculation of transport coefficients.

In absence of magnetic field, RTA and Kubo both come to the same expression on conductivity, when we establish the inverse relation ($\tau_c=1/\Gamma$) between the relaxation time $\tau_c$ of RTA and thermal width $\Gamma$ of Kubo. Interestingly, the electrical conductivity is realized as one-loop diagram of electromagnetic current-current correlator in Kubo framework, while same is realized as kinetic flow in relaxation time scale due to applied electric field in RTA framework. In presence of magnetic field, multi-component structure in transport coefficients are found and an anisotropic transportation will be observed. Using the RTA formalism~\cite{Tuchin:2011jw,Ghosh:2018cxb,Dey:2019axu,Mohanty:2018eja,Das:2019pqd,Harutyunyan:2016rxm,LIFSHITZ1981217}, it has been seen that, the transport coefficients become anisotropic with a structure of 7 viscosity components, 3 conductivity components. Along with relaxation time $\tau_c$, cyclotron time scale $\tau_B$, which is completely appeared at finite magnetic field, create different effective relaxation time scales for different components. When one approaches to calculate those multi-component transport coefficients in Kubo framework, instead of getting $\tau_B$, Landau quantization in propagation amplitude is basically appeared. For viscosity, the detailed field theoretical multi-component structures is first revealed by Ref.~\cite{Ghosh:2020wqx}, then present article is aimed for the corresponding Kubo-structure for electrical conductivity. Though Refs.~\cite{Hattori:2016lqx,Hattori:2016cnt,Fukushima:2017lvb} attempted to explore the Kubo-structure of electrical conductivity, but our calculations have found some additional information, which will be discussed in details in result section. In short, present work has zoomed in quite new information of perpendicular components of electrical conductivity, which were absent in earlier Refs.~\cite{Hattori:2016lqx,Hattori:2016cnt,Fukushima:2017lvb}. In our work we adopt the Kubo formlism equipped with the Schwinger's proper time formalism~\cite{Schwinger:1951nm,Ghosh:2020wqx,Schwartz:2013pla} to calculate electrical conductivity in the presence of an external magnetic field which involves the evaluation of correlation functions~\cite{Hosoya:1983id}.

The article is organized as follows. In Sec \ref{AB0} we study the spectral functions of complex scalar and Dirac theory and obtain their derivatives in the static limit. The results are then used to calculate conductivity in the absence of external magnetic field. The results of this section are used for the subsequent sections. In Sec~\ref{B} we compute the spectral functions of complex scalar theory and Dirac theory in the presence of magnetic field using Schwinger's proper time formalism. Here we use the propagators obtained from Schwinger's proper time formalism which basically shows that the analytic structure of the propagator contains magnetic field which has been obtained from the Lagrangian. The Landau levels are present in the propagator as summed up indices. In Sec {\ref{C}}, the conductivity is calculated in the presence and absence of external magnetic field making use of Kubo formula~\cite{Ghosh:2016yvt,Hosoya:1983id} and employing projections~\cite{Huang:2011dc} thus calculating of transport coefficients along several directions relative to the magnetic field. Various calculations used in the paper are given in the appendixes. Throughout this paper we use the natural system of units with $\hbar=c=1$ and the metric tensor $g^{\mu\nu}$ in flat space given by the signature $(+,-,-,-)$.

\section{THE SPECTRAL FUNCTION OF THE VECTOR CURRENTS} \label{AB0}
\begin{figure}[h]
	\begin{center}
		\includegraphics[angle=0,scale=0.5]{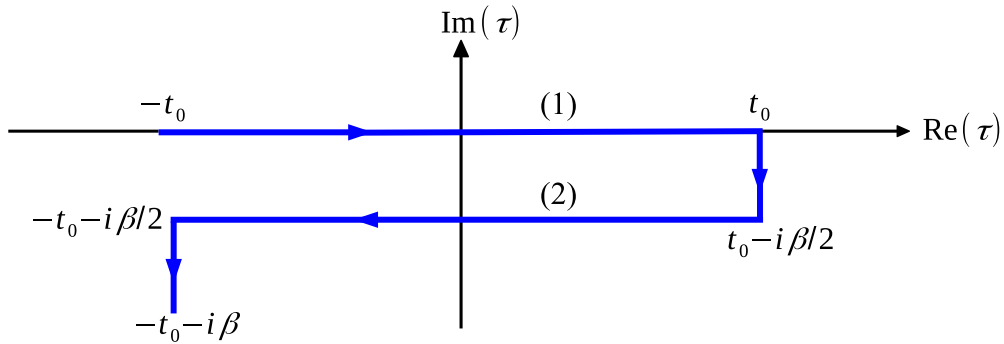}
	\end{center}
	\caption{The symmetric Schwinger-Keldysh contour $C$ in the complex time plane used in the RTF with $t_0\to\infty$ and $\beta = 1/T$. The two horizontal segments of the contour are referred to as labels `(1)' and `(2)' respectively.}
	\label{fig.contour}
\end{figure}

In the Kubo formalism, the key microscopic quantity required for the calculation of electrical conductivity tensor ($\sigma^{\mu\nu}$) is the retarded vector current correlator $\ensembleaverage{J^{\mu}(x)J^{\nu}(0}_{R}$ using which the in-medium spectral function $\rho^{\mu\nu}(q)$ can be defined as 
\begin{eqnarray}
\rho^{\mu\nu}(q) = \IM ~i \int\! d^4x e^{i q \cdot x} \ensembleaverage{J^{\mu}(x)J^{\nu}(0}_{R}.
\label{spec1}
\end{eqnarray}
Since the calculations are being done at finite temperature we have taken an ensemble average $\ensembleaverage{...}_R$ of the retarded two point correlation function. The spectral function can alternatively be expressed in terms of the time-ordered correlation function as 
\begin{eqnarray}
\rho^{\mu\nu}(q) = \tanh\FB{\frac{q^0}{2T}}\IM ~i \int\! d^4x e^{i q \cdot x} \ensembleaverage{ \mathcal{T}_cJ^{\mu}(x)J^{\nu}(0)}_{11}
\label{spectral2}
\end{eqnarray}
where $\mathcal{T}_c$ is the time-ordering operator performed on the Schwinger-Keldysh type contour $C$ on the complex time plane, as shown in Fig.~(\ref{fig.contour}); the subscript $11$ corresponds to the two points being on the real horizontal segment `(1)' of the contour $C$.

In this work, we will be mainly considering two systems: (i) system of charged scalar Bosons (spin-0) described by the complex scalar field $\phi(x)$ and 
(ii) system of charged Dirac Fermions (spin-$\frac{1}{2}$) described by the Fermion field $\psi(x)$.  The are respectively described by the following Lagrangian (densities) 
\begin{eqnarray}
&&\mathcal{L}_\text{Scalar} = \partial_{\mu}\phi^{*}\partial^{\mu}\phi - m^2\phi^{*}\phi ,\\
&&\mathcal{L}_\text{Dirac} = \bar{\psi}(i\gamma^{\mu}\partial_{\mu} - m)\psi
\end{eqnarray}
where $m$ is mass of the particles. The corresponding vector currents (Noether's current for $U(1)$ global gauge transformation) are given by 
\begin{eqnarray}
&&J^{\mu}_\text{Scalar} = ie \TB{ \phi^{*}(\partial^{\mu}\phi) - \phi^*(\partial^{\mu}\phi)}, 
\label{Current_Scalar}\\
&&J^{\mu}_\text{Dirac} = e\bar{\psi}\gamma^{\mu}\psi.
\label{Current_Dirac}
\end{eqnarray}
Using Eqs.~(\ref{Current_Scalar}) and (\ref{Current_Dirac}) the correlation functions of the vector currents $\ensembleaverage{\mathcal{T}_cJ^{\mu}(x)J^{\nu}(0)}$ for the respective type of fields considered can be calculated (see Appendix~\ref{app.1}) and we get from Eqs.~\eqref{A6} and \eqref{A11},
\begin{eqnarray}
\ensembleaverage{\mathcal{T}_cJ^{\mu}_\text{Scalar}(x)J^{\nu}_\text{Scalar}(0)}_{11} &=& -\int\!\!\!\int\!\!\frac{d^4p}{(2\pi)^4}\frac{d^4k}{(2\pi)^4}e^{-ix\cdot(p-k)}D_{11}(p;m)D_{11}(k;m) \mathcal{N}^{\mu\nu}_\text{Scalar}(k,p), \label{JJScalar} \\
\ensembleaverage{\mathcal{T}_cJ^{\mu}_\text{Dirac}(x)J^{\nu}_\text{Dirac}(0)}_{11} &=& -\int\!\!\!\int\!\!\frac{d^4p}{(2\pi)^4}\frac{d^4k}{(2\pi)^4}e^{-ix\cdot(p-k)}\tilde{D}_{11}(p;m)\tilde{D}_{11}(k;m) \mathcal{N}^{\mu\nu}_\text{Dirac}(k,p)
\label{JJDirac}
\end{eqnarray}
where $D_{11}(k;m)$ , $\tilde{D}_{11}(k;m)$ and $\mathcal{N}^{\mu\nu}_\text{Scalar,Dirac}(k,p)$ are respectively given in Eqs.~(\ref{A5}), (\ref{1012}), (\ref{A7}) and (\ref{A12}).  
Evaluating the $d^4x$ integral in the spectral function of Eq.~(\ref{spectral2}) using Eqs.~(\ref{JJScalar}) and (\ref{JJDirac}) produces a Dirac delta function $\delta^{(4)}(q-p+k)$ which is
then used to perform the $d^4p$ integral finally giving us the spectral function as 
\begin{eqnarray}
\rho^{\mu\nu}_\text{Scalar}(q) &=& -\tanh\FB{\frac{q^0}{2T}}\IM ~i\int\!\!\frac{d^4k}{(2\pi)^4}D_{11}(k;m)D_{11}(p=q+k;m)\mathcal{N}^{\mu\nu}_\text{Scalar}(k,p),
\label{spectral3s}\\
\rho^{\mu\nu}_\text{Dirac}(q) &=& -\tanh\FB{\frac{q^0}{2T}}\IM ~ i \int\!\!\frac{d^4k}{(2\pi)^4}\tilde{D}_{11}(k;m)\tilde{D}_{11}(p=q+k;m)\mathcal{N}^{\mu\nu}_\text{Dirac}(k,p).
\label{spectral3d}
\end{eqnarray}
After substituting the expressions of $D_{11}(k;m)$ and $\tilde{D}_{11}(k;m)$ from Eqs.~(\ref{A5}) and (\ref{1012}) in Eqs.~(\ref{spectral3s}) and (\ref{spectral3d}) 
we get on performing the $dk^0$ integral
\begin{eqnarray}
\rho^\munu(q) &=& \tanh\FB{\frac{q^0}{2T}}\pi\int\!\!\!\frac{d^3k}{(2\pi)^3} \frac{1}{4\omega_k\omega_p} \Big[
\SB{1+af_a(\omega_k)+af_a(\omega_p)+2f_a(\omega_k)f_a(\omega_p)} \big\{N^\munu(k^0=\omega_k)\delta(q^0-\omega_k-\omega_p) \nn \\
&& \hspace{4cm}+~ N^\munu(k^0=-\omega_k)\delta(q^0+\omega_k+\omega_p)\big\}
+ \SB{af_a(\omega_k)+af_a(\omega_p)+2f_a(\omega_k)f_a(\omega_p)} \nn \\ && \hspace{4cm}
\times  \big\{N^\munu(k^0=-\omega_k)\delta(q^0-\omega_k+\omega_p) 
+ N^\munu(k^0=\omega_k)\delta(q^0+\omega_k-\omega_p)\big\}
\Big]
\label{spectral4}
\end{eqnarray}
where 
$\omega_k = \sqrt{\vec{k}^2 + m^2}$,  $\omega_p = \sqrt{\vec{p}^2 + m^2}$ and $f_a(\omega_k)$ is the thermal distribution function given by $f_a(\omega_k) = \TB{e^{\omega_k/T} - a}^{-1}$ and  $a = \pm 1$ for scalar and Fermion particle respectively. Here we see that Eq.~(\ref{spectral4}) carries four Dirac-delta functions which belong to four different kinematic regions of the energy spectrum. They are non-zero in the kinematic regions: (i) $\sqrt{\vec{q}^2 + 4m^2} < q_0 < +\infty$ called unitary-I cut (ii) $-\infty < q_0 < -\sqrt{\vec{q}^2 + 4m^2}$ called unitary-II cut, (iii) $ \MB{q_0} < \MB{\vec{q}} $ called Landau-II cut and (iv) $ \MB{q_0} < \MB{\vec{q}} $ called Landau-I cut as they appear in Eq.~\eqref{spectral4}. For the calculation of the electrical conductivity, we will take the long wavelength limit or the static limit $\vec{q} = \vec{0}, q_0 \to 0$ of the spectral function for which case, only the Landau cuts contribute. Thus, taking the contributions from Landau cuts only, we obtain 
\begin{eqnarray}
\rho^{\mu\nu}(q_0, \vec{q}=\vec{0})&=& \tanh\FB{\frac{q^0}{2T}}\pi\!\int\!\!\!\frac{d^3k}{(2\pi)^3}\delta(q_0)\frac{1}{2\omega_k^2}
f_a(\omega_k)\{a+f_a(\omega_k)\}\{\mathcal{N}^{\mu\nu}(k_0=\omega_k) + \mathcal{N}^{\mu\nu}(k_0=-\omega_k)\}
\label{spectral5}\\
&=& \lim\limits_{\Gamma\to0}  \tanh\FB{\frac{q^0}{2T}} \!\int\!\!\!\frac{d^3k}{(2\pi)^3}\frac{\Gamma}{q_0^2 + \Gamma^2}\frac{1}{2\omega_k^2}
f_a(\omega_k)\{a+f_a(\omega_k)\} \{\mathcal{N}^{\mu\nu}(k_0=\omega_k) + \mathcal{N}^{\mu\nu}(k_0=-\omega_k)\}
\label{spectral6}
\end{eqnarray}
by making use of the Breit-Wigner representation of the Dirac-delta function. The conductivity tensor $\sigma^\munu$ is obtained using the Kubo formalism by taking the zero-momentum limit of $\rho^\munu/q^0$ or alternatively differentiating Eq.~(\ref{spectral6}) with respect to $q_0$ and taking the static limit $q^0\to0$ by means of L'Hospital's rule as 
\begin{eqnarray}
\sigma^{\mu\nu} = \FB{\frac{\partial\rho^{\mu\nu}}{\partial q_0}}_{\vec{q}=\vec{0},q^0\to0} &=& \lim\limits_{\Gamma\to0}  \frac{1}{T}\int\!\!\!\frac{d^3k}{(2\pi)^3}\frac{1}{4\omega_k^2\Gamma} f_a(\omega_k)\{a + f_a(\omega_k)\}
\Big[\mathcal{N}^{\mu\nu}(k,k)\Big|_{k_0=\omega_k} + \mathcal{N}^{\mu\nu}(k,k)\Big|_{k_0=-\omega_k}\Big]
\label{spectral7}
\end{eqnarray}
where, the expressions of $\mathcal{N}^{\mu\nu}_\text{Scalar,Dirac}(k,k)$ can be read off respectively from Eqs.~(\ref{A8}) and (\ref{A13}) as
\begin{eqnarray}
\mathcal{N}^{\mu\nu}_\text{Scalar}(k,k) &=& 4e^2k^{\mu}k^{\nu},
\label{NScalar}\\
\mathcal{N}^{\mu\nu}_\text{Dirac}(k,k) &=& -4e^2\Big[2k^{\mu}k^{\nu} - g^{\mu\nu}(k^2-m^2)\Big].
\label{NDirac}
\end{eqnarray}
In Eq.~(\ref{spectral6}), we have introduced a parameter $\Gamma$ to get a finite value of $\rho^{\mu\nu}(q)$ in the $\vec{q}, q_0 \to 0$ limit. Here $\Gamma$ stands for the thermal width which basically tells us that interactions are present in the medium and this causes dissipation. From Eq.~(\ref{spectral7}) it is seen that transport coefficients are inversely proportional to $\Gamma$.

\section{VECTOR CURRENT SPECTRAL FUNCTION IN THE PRESENCE OF MAGNETIC FIELD}
\label{B}
In the presence of an external magnetic field $\vec{B}$ characterized by the four-potential $A^\mu_\text{ext}(x)$, the corresponding Lagrangians for the scalar and Dirac field are modified to 
\begin{eqnarray}
\mathcal{L}_\text{Scalar} &=& D^{*\mu}\phi^\dagger D_{\mu}\phi - m^2\phi^\dagger\phi,
\label{ScalarMag}\\
\mathcal{L}_\text{Dirac} &=& \psibar\FB{i\gamma^{\mu}D_{\mu} - m}\psi
\label{DiracMag}
\end{eqnarray}
where $D^{\mu} = \partial^{\mu} + ie A^{\mu}_\text{ext}$ is the covariant derivative and $e>0$ being the electric charge of the particle. The vector currents in the presence of magnetic field are then given by  
\begin{eqnarray}
&&J^{\mu}_\text{Scalar} = ie\TB{\phi^{\dagger}(D^{\mu}\phi) - (D^{*\mu}\phi^{\dagger})\phi},
\label{JScalarB}\\
&&J^{\mu}_\text{Dirac} = e\overline{\psi}\gamma^{\mu}\psi.
\label{JDiracB}
\end{eqnarray}
Considering the external magnetic field $\vec{B}$ in the $\hat{z}$-direction, we calculate the vector current correlation function $\ensembleaverage{\mcTc J^\mu(x)J^\nu(0)}^B_{11}$ for both the scalar and Dirac cases (see Appendix~\ref{app.B}) and we get from Eqs.~(\ref{B12}) and (\ref{B18})
\begin{eqnarray}
\ensembleaverage{\mathcal{T}_cJ^{\mu}_\text{Scalar}(x)J^{\nu}_\text{Scalar}(0)}_{11}^{B} &=& - \int\!\!\!\int\!\!\frac{d^4p}{(2\pi)^4}\frac{d^4k}{(2\pi)^4}e^{-ix\cdot(p-k)}
\sum_{l=0}^{\infty}\sum_{n=0}^{\infty}D_{11}(p_{11};m_n)D_{11}(k_{11};m_l)\mathcal{N}_{ln;\text{Scalar}}^{\mu\nu}(k,p),
\label{CorrScaB}\\
\ensembleaverage{\mathcal{T}_cJ^{\mu}_\text{Dirac}(x)J^{\nu}_\text{Dirac}(0)}_{11}^{B} &=& - \int\!\!\!\int\!\!\frac{d^4p}{(2\pi)^4}\frac{d^4k}{(2\pi)^4}e^{-ix\cdot(p-k)}
\sum_{l=0}^{\infty}\sum_{n=0}^{\infty}\tilde{D}_{11}(p_{11};m_n)\tilde{D}_{11}(k_{11};m_l)\mathcal{N}_{ln;\text{Dirac}}^{\mu\nu}(k,p)
\label{CorrDirB}
\end{eqnarray}
where $D_{11}$, $\tilde{D}_{11}$ and $\mathcal{N}_{ln;\text{Scalar,Dirac}}^{\mu\nu}$ can be read off from Eqs.~(\ref{A5}), (\ref{1012}), (\ref{eq.N.Scalar}) and (\ref{B16}) respectively. In the above equation, 
\begin{eqnarray}
m_l = \sqrt{m^2  + (2l+1-2s)eB},
\end{eqnarray}
which are the `\textit{effective Landau level dependent}' masses in the presence of external magnetic field and $s$ corresponds to the spin of the particle being considered: 
\begin{eqnarray}
s = 
\begin{cases}
0 ~~~\text{for Scalar Boson}, \nn \\
  1/2 ~~~\text{for Dirac Fermion.}
 \end{cases}
\end{eqnarray}

Proceeding in the same manner as we did in the 
absence of magnetic field, we calculate the vector current spectral function by performing the $d^4x$ integral of Eq.~\eqref{spectral2} after substituting Eqs.~(\ref{CorrScaB}) and (\ref{CorrDirB}) into it and obtain a Dirac delta function $\delta^4(q-p+k)$. The $d^4p$ integral is then performed using the delta function yielding
\begin{eqnarray}
&&\rho^{\mu\nu}_\text{Scalar}(q) = -\tanh\FB{\frac{q^0}{2T}}\IM~i \int\!\!\!\frac{d^4k}{(2\pi)^4}\sum_{n=0}^\infty\sum_{l=0}^\infty
D_{11}(k_\parallel;m_l)D_{11}(p_\parallel = q_\parallel +k_\parallel;m_n)\mathcal{N}^{\mu\nu}_{ln;\text{Scalar}}(k,p=q+k),
\label{rhoBS}\\
&&\rho^{\mu\nu}_\text{Dirac}(q) = -\tanh\FB{\frac{q^0}{2T}}\IM~i \int\!\!\!\frac{d^4k}{(2\pi)^4}\sum_{n=0}^{\infty}\sum_{l=0}^{\infty}
\tilde{D}_{11}(k_\parallel;m_l)\tilde{D}_{11}(p_\parallel = q_\parallel +k_\parallel;m_n)\mathcal{N}^{\mu\nu}_{ln;\text{Dirac}}(k,p=q+k).
\label{rhoBD}
\end{eqnarray}
Using the expressions of $D_{11}$ and $\tilde{D}_{11}$ from Eqs.~(\ref{A5}) and (\ref{1012}) and performing the $dk_0$ integral, we get
\begin{eqnarray}
\rho^{\mu\nu}(q) &=& \tanh\FB{\frac{q^0}{2T}}\pi \sum_{l=0}^{\infty}\sum_{n=0}^{\infty}\int\!\!\!\frac{d^3k}{(2\pi)^3}\frac{1}{4\omega_{kl}\omega_{pn}}
\Big[\big\{1 + af_a(\omega_{kl}) + af_a(\omega_{pn}) + 2f_a(\omega_{kl})f_a(\omega_{pn})\big\} \nn \\
&& \hspace{-1cm}\times \big\{\mathcal{N}^{\mu\nu}_{ln}(k_0 = -\omega_{kl})\delta(q_0+\omega_{kl}+\omega_{pn}) 
+\mathcal{N}^{\mu\nu}_{ln}(k_0=\omega_{kl})\delta(q_0-\omega_{kl}-\omega_{pn})  \big\}
  + \big\{af_a(\omega_{kl}) + af_a(\omega_{pn}) + 2f_a(\omega_{kl})f_a(\omega_{pn})\big\} \nn \\
&&  \times \big\{\mathcal{N}_{ln}^{\mu\nu}(k_0=-\omega_{kl})\delta(q_0-\omega_{kl}+\omega_{pn})+
\mathcal{N}^{\mu\nu}_{ln}(k_0=\omega_{kl})\delta(q_0 + \omega_{kl} -\omega_{pn})\big\}\Big]
\label{rhoB}
\end{eqnarray}
where $\omega_{kl}= \sqrt{k_z^2+m_l^2}$ and $\omega_{pn} = \sqrt{(p_z+q_z)^2 + m_n^2}$.
We will now take the long wavelength limit of Eq.~(\ref{rhoB}) so that only the Landau cuts contribute and we are left with
\begin{eqnarray}
\rho^{\mu\nu}( q_0, \vec{q}=\vec{0}) &=& \tanh\FB{\frac{q^0}{2T}}\pi\sum_{l=0}^{\infty}\sum_{n=0}^{\infty}\int\!\!\!\frac{d^3k}{(2\pi)^3}\frac{1}{4\omega_{kl}\omega_{kn}}
\big\{af_a(\omega_{kl}) + af_a(\omega_{kn}) + 2f_a(\omega_{kl})f_a(\omega_{kn}) \big\} \nn \\
&& \times~ \big\{\mathcal{N}^{\mu\nu}_{ln}(k_0 = -\omega_{kl})\delta(q_0-\omega_{kl} + \omega_{kn}) + \mathcal{N}^{\mu\nu}_{ln}(k_0=\omega_{kl})\delta(q_0+\omega_{kl}-\omega_{kn}) \big\}\nn \\
&=& \lim\limits_{\Gamma\to0} \tanh\FB{\frac{q^0}{2T}} \sum_{l=0}^{\infty}\sum_{n=0}^{\infty}\int\!\!\!\frac{d^3k}{(2\pi)^3}\frac{1}{4\omega_{kl}\omega_{kn}}
\big\{af_a(\omega_{kl}) + af_a(\omega_{kn}) + 2f_a(\omega_{kl})f_a(\omega_{kn}) \big\} \nn \\
&& \times \Big[\mathcal{N}^{\mu\nu}_{ln}(k_0=-\omega_{kl})\frac{\Gamma}{\Gamma^2+(q_0 - \omega_{kl} + \omega_{kn})^2}   +\mathcal{N}^{\mu\nu}_{ln}(k_0 = \omega_{kl})\frac{\Gamma}{\Gamma^2 + (q_0+\omega_{kl}-\omega_{kn})^2}\Big]
\label{rhoB0}
\end{eqnarray}
where we have again introduced the Breit-Wigner form of the Dirac-delta function. 
Proceeding in the same manner as the zero magnetic field case, we differentiate Eq.~(\ref{rhoB0}) with respect to $q^0$ and take the limit $q^0\to0$ to obtain the conductivity tensor at $B\ne 0$ as
\begin{eqnarray}
\sigma^{\mu\nu}_B = \frac{\partial\rho^{\mu\nu}}{\partial q_0}\Big|_{\vec{q}\to \vec{0},q_0 = 0} 
&=& \lim\limits_{\Gamma\to 0} \sum_{l=0}^{\infty}\sum_{n=0}^{\infty}\frac{1}{2T}\int\!\!\!\frac{d^3k}{(2\pi)^3}\frac{1}{4\omega_{kl}\omega_{kn}}
\frac{\Gamma}{\Gamma^2 + (\omega_{kl} -\omega_{kn})^2} \nn \\
&& \times~ \big\{af_a(\omega_{kl}) + af_a(\omega_{kn}) + 2f_a(\omega_{kl})f_a(\omega_{kn})\big\} 
 \Big[\mathcal{N}^{\mu\nu}_{ln}(k,k)\big|_{k_0 = \omega_{kl}} + \mathcal{N}^{\mu\nu}_{ln}(k,k)\big|_{k_0 = -\omega_{kl}}\Big]
\label{rhoBTrans}
\end{eqnarray}
where the explicit expressions of $\mathcal{N}^{\mu\nu}_{ln;\text{Scalar,Dirac}}(k,k)$ are given in Eqs.~(\ref{B13}) and (\ref{B18})
 as
\begin{eqnarray}
\mathcal{N}_{ln;\text{Scalar}}^{\mu\nu}(k,k) &=& 16e^2\mathcal{A}_{ln}(\kper^2)k^\mu k^\nu, \label{NBScalar} \\
\mathcal{N}^{\mu\nu}_{ln;\text{Dirac}}(k,k) &=& -8e^2\Big[8\FB{2k_{\perp}^{\mu}k_{\perp}^{\nu} - k_{\perp}^2g^{\mu\nu}}\mathcal{B}_{ln}(\kper^2)  + \SB{ 2k_\parallel^{\mu}k_\parallel^{\nu}-\gpll^{\mu\nu}(k_\parallel^2-m^2)}\mathcal{C}_{ln}(\kper^2)\nn \\
&& + g_{\perp}^{\mu\nu}\FB{k_\parallel^2-m^2}\mathcal{D}_{ln}(\kper^2)+ 2\FB{k_\parallel^{\nu}k_{\perp}^{\mu} + k_\parallel^{\mu}k_{\perp}^{\nu}}\mathcal{E}_{ln}(\kper^2)\Big]. \label{NBDirac1} 
\end{eqnarray}
where the functions $\mathcal{A}_{ln}(\kper^2)$, $\mathcal{B}_{ln}(\kper^2)$ $\mathcal{c}_{ln}$, ... involving the Laguerre polynomials are provided in Eqs.~\eqref{eq.Anl} and \eqref{eq.Bnl}-\eqref{eq.Enl}.


\section{ELECTRICAL CONDUCTIVITY FROM SPECTRAL FUNCTION IN KUBO FORMALISM} \label{C}
The electrical conductivity ($\sigma$) for a relativistic  fluid can be calculated from the in-medium vector current spectral functions using the Kubo relation~\cite{Hattori:2016cnt}. Starting with the $B=0$ case, the isotropic electrical conductivity $\sigma$ is related to the conductivity tensor $\sigma^\munu$ via
\begin{eqnarray}
\sigma = \mathcal{P}_\munu \sigma^\munu \label{eq.sigma}
\end{eqnarray}
where the projector
\begin{eqnarray}
\mathcal{P}_\munu = -\frac{1}{3}g_{\alpha\beta}\Delta^{\alpha\mu}\Delta^{\beta\nu} \label{eq.proj}
\end{eqnarray}
in which $\Delta^{\mu\nu} = g^{\mu\nu}-u^{\mu}u^{\nu}$. Substituting Eq.~\eqref{spectral7} into Eq.~\eqref{eq.sigma}, we arrive at the following expression of electrical conductivity in absence of external magnetic field
\begin{eqnarray}
\sigma = \frac{1}{T}\int\!\!\!\frac{d^3k}{(2\pi)^3}  \frac{1}{4\omega_k^2\Gamma} f_a(\omega_k)\SB{a + f_a(\omega_k)} \mathcal{N}_\text{Scalar,Dirac}(\vec{k})
\label{condB0}
\end{eqnarray}
where, 
\begin{eqnarray}
\mcN(\vec{k}) = \frac{1}{2}\mathcal{P}_\munu 
\SB{\mathcal{N}^\munu(k,k)\big|_{k^0=\omega_k} + \mathcal{N}^\munu(k,k)\big|_{k^0=-\omega_k}}.
\label{Nmunu}
\end{eqnarray}
On substitution of Eqs.~(\ref{NScalar}), (\ref{NDirac}) and \eqref{eq.proj} into Eq.~(\ref{Nmunu}) and bit simplification yields, 
\begin{eqnarray}
\mathcal{N}_\text{Scalar} = \frac{4}{3}e^2\vec{k}^2
\label{NSc} ~~\text{and}~~
\mathcal{N}_\text{Dirac} = -\frac{8}{3}e^2\vec{k}^2.
\label{NDi}
\end{eqnarray}

Using Eq.~(\ref{NSc}) into Eq.~(\ref{condB0}), we obtain the following well known expression of conductivities at $B=0$
\begin{eqnarray} 
\sigma_\text{Scalar} &=& \frac{2e^2}{3T}\int\!\!\!\frac{d^3k}{(2\pi)^3}\frac{\vec{k}^2}{\omega_k^2\Gamma}f_+(\omega_k)\SB{1+f_+(\omega_k)},
\label{ConB0Sc}\\
\sigma_\text{Dirac} &=& \frac{4e^2}{3T}\int\!\!\!\frac{d^3k}{(2\pi)^3}\frac{\vec{k}^2}{\omega_k^2\Gamma}f_-(\omega_k)\SB{1-f_-(\omega_k)}~.
\label{ConB0SDir}
\end{eqnarray}

In the presence of an external magnetic field, $\sigma^{\mu\nu}_B$ is not symmetric as in $B=0$ case. Thus the different components of conductivity are obtained by application of the corresponding projectors on the conductivity tensor $\sigma^{\mu\nu}$. In RTA~\cite{Dey:2019axu,Mohanty:2018eja}, it is found that 
electrical conductivity breaks up into multiple components with respect to the direction of the external magnetic field. In our Kubo formalism, the different components of conductivities are obtain from
\begin{eqnarray}
\sigma^\upsilon = \mathcal{P}_{\mu\nu}^\upsilon\sigma^{\mu\nu}_B ~~;~~ \upsilon \in \{\parallel, \perp, \times \}
\label{proj1}
\end{eqnarray}
where, the projectors $\mathcal{P}_{\mu\nu}^\upsilon$ are given by 
\begin{eqnarray}
\mathcal{P}_{\mu\nu}^\parallel &=& b^{\alpha}b^{\beta}\Delta_{\alpha\mu}\Delta_{\beta\nu},
\label{projparall}\\
\mathcal{P}_{\mu\nu}^\perp &=& -\frac{1}{2}\Xi^{\alpha\beta}\Delta_{\alpha\mu}\Delta_{\beta\nu},
\label{projperp}\\
\mathcal{P}_{\mu\nu}^\times &=& \frac{1}{2}b^{\alpha\beta}\Delta_{\alpha\mu}\Delta_{\beta\nu}
\label{projcross}
\end{eqnarray}
where $b^\mu = \frac{1}{2B}\varepsilon^{\mu\nu\alpha\beta}F_{\nu\alpha}u_{\beta}$, $F_{\mu\nu} = \FB{\partial_{\mu}A_{\nu,\text{ext}} - \partial_{\nu}A_{\mu,\text{ext}}}$
is the field strength tensor, anti-symmetric in Lorentz indices, $b^{\mu\nu}= \varepsilon^{\mu\nu\alpha\beta}b_{\alpha}u_{\beta}$ and 
$\Xi^{\mu\nu} = \Delta^{\mu\nu} + b^{\mu}b^{\nu}$. 
In the LRF of the bath, $b^{\mu}_\text{LRF} = (0,0,0,1)$ points along the direction of external magnetic field.

Substituting Eqs.~\eqref{projparall}-\eqref{projcross} into Eq.~\eqref{proj1}, we can explicitly express the conductivity coefficients $\sigma^{\pll,\perp,\times}$ in terms of the different component of conductivity tensor as
\begin{eqnarray}
\sigma^\perp &=& \frac{1}{2}\FB{\sigma^{11}+\sigma^{22}} \label{eq.perp}\\
\sigma^\parallel &=& \sigma^{33} \\
\sigma^\times &=& \frac{1}{2}\FB{\sigma^{12}-\sigma^{21}}.
\label{eq.hall}
\end{eqnarray}
From Eqs.~\eqref{eq.perp}-\eqref{eq.hall}, we can identify $\sigma^\perp = \sigma^{11} = \sigma_0 $ as the transverse conductivity, $\sigma^\parallel = \sigma^{33} = \sigma_0+\sigma_2$ as the longitudinal conductivity, and, $\sigma^\times = \sigma^{12} = \sigma_1$ as the Hall conductivity, where $\sigma_0$, $\sigma_1$ and $\sigma_2$ are the notations used in Refs.~\cite{Harutyunyan:2016rxm,Harutyunyan:2018mpe,Dash:2020vxk}.

In the presence of external magnetic field, the different conductivity components are obtained by substituting Eq.~\eqref{rhoBTrans} into Eq.~\eqref{proj1} as
\begin{eqnarray}
\sigma^\upsilon_B = \frac{1}{T}\sum_{n=0}^{\infty}\sum_{l=0}^{\infty}\int\!\!\!\frac{d^3k}{(2\pi)^3}\frac{1}{4\omega_{kl}\omega_{kn}}\frac{\Gamma}{(\omega_{kl}-\omega_{kn})^2 + \Gamma^2}
\big\{af_a(\omega_{kl}) + af_a(\omega_{kn}) + 2f_a(\omega_{kl})f_a(\omega_{kn}) \big\}\mathcal{N}_{ln}^\upsilon(\vec{k})
\label{condB1}
\end{eqnarray}
where
\begin{eqnarray}
\mathcal{N}_{ln}^\upsilon = \frac{1}{2}\mathcal{P}_{\mu\nu}^\upsilon\Big\{\mathcal{N}_{ln}^{\mu\nu}(k,k)\big|_{k_0=\omega_{kl}} + 
\mathcal{N}_{ln}^{\mu\nu}(k,k)\big|_{k_0=-\omega_{kl}}  \Big\}.
\label{NlnB}
\end{eqnarray}
Let us now substitute Eqs.~(\ref{NBScalar}), (\ref{NBDirac1}) and (\ref{projparall})-\eqref{projcross} into Eq.~\eqref{NlnB} to get,
\begin{eqnarray}
\mathcal{N}_{ln;\text{Scalar}}^\parallel(\vec{k}) &=& 16e^2\mathcal{A}_{ln}(\kper^2)k_z^2,
\label{N11}\\
\mathcal{N}_{ln;\text{Scalar}}^\perp(\vec{k}) &=& -8e^2\mathcal{A}_{ln}(\kper^2)k_{\perp}^2,
\label{N2}\\
\mathcal{N}_{ln;\text{Dirac}}^\parallel(\vec{k}) &=& -8e^2\Big\{8k_{\perp}^2\mathcal{B}_{ln}(\kper^2) + (k_z^2+\omega_{kl}^2 -m^2)\mathcal{C}_{ln}(\kper^2) \Big\},
\label{N3}\\
\mathcal{N}_{ln;\text{Dirac}}^\perp(\vec{k}) &=& -8e^2\Big\{k_z^2 + m^2 -\omega_{kl}^2\Big\}\mathcal{D}_{ln}(\kper^2),
\label{N4} \\
\mathcal{N}_{ln;\text{Scalar}}^\times(\vec{k}) &=& \mathcal{N}_{ln;\text{Dirac}}^\times(\vec{k}) =0.
\end{eqnarray}

During the entire calculation, we have introduced the thermal width $\Gamma$ as a parameter, although it can be calculated microscopically from the interaction Lagrangian. In most general cases, $\Gamma$ will depend on temperature ($T$), magnetic field ($B$) as well as momentum $\vk$ \textit{i.e.} $\Gamma = \Gamma (T,B,\vk)$. However, taking appropriate momentum average we can approximate $\Gamma$ to be independent of momentum and thus take it outside the $d^2\kper$ integral of Eq.~\eqref{condB1}. This will enable us to perform the analytic $d^2\kper$ integral of Eq.~\eqref{condB1} and we get, the following simplified expressions of the electrical conductivities in presence of constant external magnetic field:
\begin{eqnarray}
\sigma^\upsilon_B  = \frac{1}{T} \sum_{n=0}^{\infty}\sum_{l=0}^{\infty}\int_{-\infty}^{+\infty}\frac{dk_z}{2\pi}\frac{1}{4\omega_{kl}\omega_{kn}} 
\frac{\Gamma}{(\omega_{kl}-\omega_{kn})^2 + \Gamma^2}
\big\{af_a(\omega_{kl}) + af_a(\omega_{kn} )+ 2f_a(\omega_{kl})f_a(\omega_{kn})\big\}\mathcal{\tilde{N}}_{ln}^\upsilon(k_z)
\label{condgeneral}
\end{eqnarray}
where, 
\begin{eqnarray}
\mathcal{\tilde{N}}_{ln}^\upsilon(k_z) = \int\!\!\frac{d^2k_{\perp}}{(2\pi)^2}\mathcal{N}_{ln}^\upsilon(\vk). 
\label{Ntilde}
\end{eqnarray}
Using Eqs.~(\ref{N11})-(\ref{N4}) in Eq.~(\ref{Ntilde}) we get
\begin{eqnarray}
\mathcal{\tilde{N}}_{ln;\text{Scalar}}^\parallel(k_z) &=& 16e^2k_z^2\mathcal{A}_{ln}^{(0)},
\label{Ntilde1}\\
\mathcal{\tilde{N}}_{ln;\text{Scalar}}^\perp(k_z) &=& -8e^2\mathcal{A}_{ln}^{(2)},
\label{Ntilde2}\\
\mathcal{\tilde{N}}_{ln;\text{Dirac}}^\parallel(k_z) &=& -8e^2\Big[8\mathcal{B}_{ln}^{(2)} + \mathcal{C}_{ln}^{(0)}\FB{\omega_{kl}^2 + k_z^2 -m^2}\Big],
\label{Ntilde3}\\
\mathcal{\tilde{N}}_{ln;\text{Dirac}}^\perp(k_z) &=& -8e^2\mathcal{D}_{ln}^{(0)}\FB{k_z^2 + m^2 -\omega_{kl}^2},
\label{Ntilde4}
\\
\mathcal{\tilde{N}}_{ln;\text{Scalar}}^\times(k_z) &=& \mathcal{\tilde{N}}_{ln;\text{Dirac}}^\times(k_z) =0
\end{eqnarray}
where\begin{eqnarray}
\mathcal{A}_{ln}^{(j)} &=& \int\!\! \frac{d^2\kper}{(2\pi)^2}\mathcal{A}_{ln}(\kper^2) \FB{\kper^2}^{j/2}, \label{eq.Aln.j}\\
\mathcal{B}_{ln}^{(j)} &=& \int\!\! \frac{d^2\kper}{(2\pi)^2}\mathcal{B}_{ln}(\kper^2) \FB{\kper^2}^{j/2}, \label{eq.Bln.j} \\
\mathcal{C}_{ln}^{(j)} &=& \int\!\! \frac{d^2\kper}{(2\pi)^2}\mathcal{C}_{ln}(\kper^2) \FB{\kper^2}^{j/2}, \label{eq.Cln.j} \\
\mathcal{D}_{ln}^{(j)} &=& \int\!\! \frac{d^2\kper}{(2\pi)^2}\mathcal{D}_{ln}(\kper^2) \FB{\kper^2}^{j/2}. \label{eq.Dln.j} 
\end{eqnarray}
Making use of the orthogonality of the Laguerre polynomials present in the functions $\mathcal{A}_{ln}(\kper^2)$, $\mathcal{B}_{ln}(\kper^2)$, $\cdots$,  $\mathcal{D}_{ln}(\kper^2)$, 
the $d^2\kper$ integrals of Eqs.~\eqref{eq.Aln.j}-\eqref{eq.Dln.j} can now be performed and the analytic expressions of the quantities $\mathcal{A}^{(j)}_{ln}$, $\mathcal{B}^{(j)}_{ln}$, $\cdots$,  $\mathcal{D}^{(j)}_{ln}$ are provided in Appendix~\ref{app.c}. Eqs.~(\ref{eq.Z0}-\ref{eq.Z1}) carry very important information, which will be discussed in result section. The Kronecker delta functions will impose restrictions on Landau levels of propagators, which finally fix the allowed (quantum) fluctuations for electric charge transportation.


\section{NUMERICAL RESULTS AND DISCUSSIONS} \label{sec.numerics}
Here, we will explore on numerical results of Kubo expressions, where we will eventually notice a rich quantum field theoretical information in transport properties of relativistic fluid in presence of magnetic field. To realize the field theoretical modification part, we will take a quick revisit of kinetic theory based expressions and then we will find out step by step changes. 

In absence of magnetic field, kinetic theory based expression of conductivity becomes exactly same as the one-loop Kubo expression, obtained in Eqs.~(\ref{ConB0Sc}) and (\ref{ConB0SDir}) for Scalar and Dirac particle respectively. Relaxation time approximation (RTA) of kinetic theory provide the form of conductivity in terms of relaxation time $\tau_c$ as~\cite{Dey:2019axu,Mohanty:2018eja,LIFSHITZ1981217,Harutyunyan:2016rxm}
\begin{eqnarray}
\sigma_\text{Scalar,Dirac} = \frac{ge^2}{3T}\!\int\!\!\!\frac{d^3k}{(2\pi)^3}\frac{\vec{k}^2\tau_c}{\omega_k^2}f_a(\omega_k)\SB{1+ af_a(\omega_k)}~.
\label{RTA_B0}
\end{eqnarray}
Realizing inverse relation between thermal width ($\Gamma$) and relaxation time ($\tau_c$) as $\tau_c=1/\Gamma$ and spin degeneracy factor $g=2,~4$ for charged scalar and Dirac particle, one can see that RTA expressions~(\ref{RTA_B0}) and Kubo expressions~(\ref{ConB0Sc}), (\ref{ConB0SDir}) are exactly same.

Now in presence of magnetic field, the expression of electrical conductivity components in kinetic theory approach are addressed in 
Refs.~\cite{Ghosh:2018cxb,Dey:2019axu,Dey:2019vkn,Dash:2020vxk,LIFSHITZ1981217,Harutyunyan:2016rxm,Feng:2017tsh}:
\begin{eqnarray}
\sigma^\upsilon  =  \frac{ge^2}{3T}\!\int\!\!\!\frac{d^3k}{(2\pi)^3}\frac{\vec{k}^2}{\omega_k^2}\tau^\upsilon f_a(\omega_k) \SB{1+a f_a(\omega_k)}~,
\label{cond_RTAB}
\end{eqnarray}
where effective relaxation time $\tau^\upsilon$ can be expressed in terms of (collisional) relaxation time $\tau_c$ and inverse time scale of cyclotron frequency $\tau_B=\om_B^{-1}=\frac{\om_k}{eB}$ as
\begin{eqnarray}
\tau^{\parallel} &=&\tau_c,
\nn\\
\tau^{\perp} &=&\frac{\tau_c}{1+(\frac{\tau_c}{\tau_B})^2},
\nn\\
\tau^{\times} &=&\frac{\tau_c(\frac{\tau_c}{\tau_B})}{1+(\frac{\tau_c}{\tau_B})^2}~.
\end{eqnarray}

Few comments on Hall conductivity are in order here. If we analyze the RTA-based Hall expression, then effective relaxation times $\tau^{\times}$ for particle and anti-particle will be opposite in sign, which represent opposite Hall current directions of them. So, in RTA, a finite Hall conductivity is expected. However, in one-loop Kubo formalism, the Hall conductivity is coming zero as we explain in the following. From Eq.~\eqref{eq.hall}, we see that, a non-zero value of $\sigma^\times$ is possible iff $\sigma^\munu$ has some anti-symmetric term. In Eqs.~(\ref{NBScalar}) and (\ref{NBDirac1}), we see $\mathcal{N}^\munu_{ln}$ are symmetric in the Lorentz indices and so the electromagnetic spectral function in Eq.~\eqref{rhoBTrans} i.e. $\sigma^\munu = \sigma^\numu$. Thus, in the one-loop Kubo formalism, the Hall conductivity becomes zero. The symmetry of $\sigma^\munu$ can be explained in a more general fashion as follows. The electromagnetic spectral functions for the system of charged scalar (Dirac) particles defined in Eq.~(\ref{spec1}) can alternatively thought as the imaginary part of the one-loop photon self energy in scalar-QED. In Ref.~\cite{Nieves:1988qz}, it has been shown that, the photon self energy can have an anti-symmetric term iff the $P$ or $CP$-odd effects are present in the Lagrangian or in the background medium. The vanishing of Hall type transport coefficients is also observed in earlier calculations for shear viscosity~\cite{Ghosh:2020wqx,Finazzo:2016mhm,Critelli:2014kra}. Therefore, our focal components are parallel and perpendicular conductivity for numerical discussion.

Now if we impose Landau quantization in final expression (\ref{cond_RTAB}) then we can get their quantum version expressions.
The main modification will occur in the dispersion relation of energy $\om_k$ and phase space $\int d^3k$ by the following replacements:
\begin{eqnarray}
\om_k = \sqrt{\vk^2+m^2} &~~~\rightarrow~~~&  \omega_{kl} = \sqrt{k_z^2+m^2+(2l+1-2s)eB} \\
g\int \!\!\!\frac{d^3k}{(2\pi)^3} &~~~\rightarrow~~~&  
\begin{cases}
\sum_{l=0}^\infty 2\frac{|{ e}|B}{2\pi} \int\limits^{+\infty}_{-\infty} \frac{dk_z}{2\pi} ~~~~ \text{(for Scalar particle)},\\
\sum_{l=0}^\infty 2g_l \frac{|{ e}|B}{2\pi} \int\limits^{+\infty}_{-\infty} \frac{dk_z}{2\pi} ~~~~ \text{(for Dirac particle)},
\end{cases}
\label{CM_QM}
\end{eqnarray}
with $g_l = \FB{2 - \delta_{l,0}}$ being the Landau level dependent spin degeneracy (implying that the lowest Landau level is spin non-degenerate). The momentum quantization in perpendicular direction can be assumed roughly as 
\begin{eqnarray} 
k_x^2\approx k_y^2\approx \frac{1}{2}\FB{k_x^2+k_y^2}&=& 
\begin{cases}
\FB{l+1/2}eB ~~~~ \text{(for Scalar particle)}, \\
 l{ e}B ~~~~ \text{(for Dirac particle)}.
\end{cases}
\end{eqnarray} 
Imposing the above quantization in Eq.~(\ref{cond_RTAB}), we will get
\begin{eqnarray}
\sigma^{\perp}_\text{Scalar} &=& \frac{2e^2}{T} \sum_{l=0}^\infty \frac{|{ e}|B}{2\pi} 
\int\limits^{+\infty}_{-\infty} \frac{dk_z}{2\pi} \frac{{(l+\frac{1}{2})|{ e}|B}}{\om^2_{l}} \tau^{\perp}
f_+(\om_{kl})\SB{1+f_+(\om_{kl})},  \label{eq.QM}
\\
\sigma^{\perp}_\text{Dirac} &=& \frac{2e^2}{T} \sum_{l=0}^\infty g_l \frac{|{ e}|B}{2\pi} 
\int\limits^{+\infty}_{-\infty} \frac{dk_z}{2\pi} \frac{{l|{ e}|B}}{\om^2_{l}} \tau^{\perp}
f_-(\om_{kl})\SB{1-f_-(\om_{kl})},
\\
\sigma^{\parallel}_\text{Scalar} &=& \frac{2e^2}{T}  \sum_{l=0}^\infty \frac{|{ e}|B}{2\pi} 
\int\limits^{+\infty}_{-\infty} \frac{dk_z}{2\pi} \frac{k_z^2}{\om^2_{l}} \tau^{\parallel}
f_+(\om_{kl})\SB{1+f_+(\om_{kl})}, \label{eq.scalar.qm}
\\
\sigma^{\parallel}_\text{Dirac}&=& \frac{2e^2}{T}  \sum_{l=0}^\infty g_l \frac{|{ e}|B}{2\pi} 
\int\limits^{+\infty}_{-\infty} \frac{dk_z}{2\pi} \frac{k_z^2}{\om^2_{l}} \tau^{\parallel}
f_-(\om_{kl})\SB{1-f_-(\om_{kl})}. 
\label{Lsig_QM}
\end{eqnarray}
For our convenient, let us call $\sigma^{\parallel,\perp}$ from Eq.~(\ref{cond_RTAB}) as classical mechanical (CM) expressions (say $\sigma^{\parallel,\perp}_\text{CM}$); from Eqs.~\eqref{eq.QM}-(\ref{Lsig_QM}) as quantum mechanical (QM) expressions (say $\sigma^{\parallel,\perp}_\text{QM}$); from Eq.~(\ref{condgeneral}) as quantum field theoretical (QFT) expressions (say $\sigma^{\parallel,\perp}_\text{QFT}$). We should accept that their naming are little bit of misleading, e.g. CM expressions carry FD/BE distributions (quantum aspects of statistical mechanics), but they might be allowed for description purpose. Similarly in QM expressions, we have used Landau quantization but it carries classical information of cyclotron motion with time period $\tau_B$. Due to Landau quantization of energy, we can assume different quantized circular orbit. This QM picture might be compared with the Bohr's semi-classical quantized orbital motion of electron in hydrogen atom, which lacking of detail probabilistic picture of electron, obtained after solving its Schrodinger's equation. We believe that our QFT expressions carry that detail quantum mechanical probabilistic picture as well as proper relativistic impositions. Generating curves of CM, QM and QFT expressions, we have discussed below and tried to interpret them one by one.

\begin{figure} [h]
	\begin{center}
		\includegraphics[angle=-90,scale=0.30]{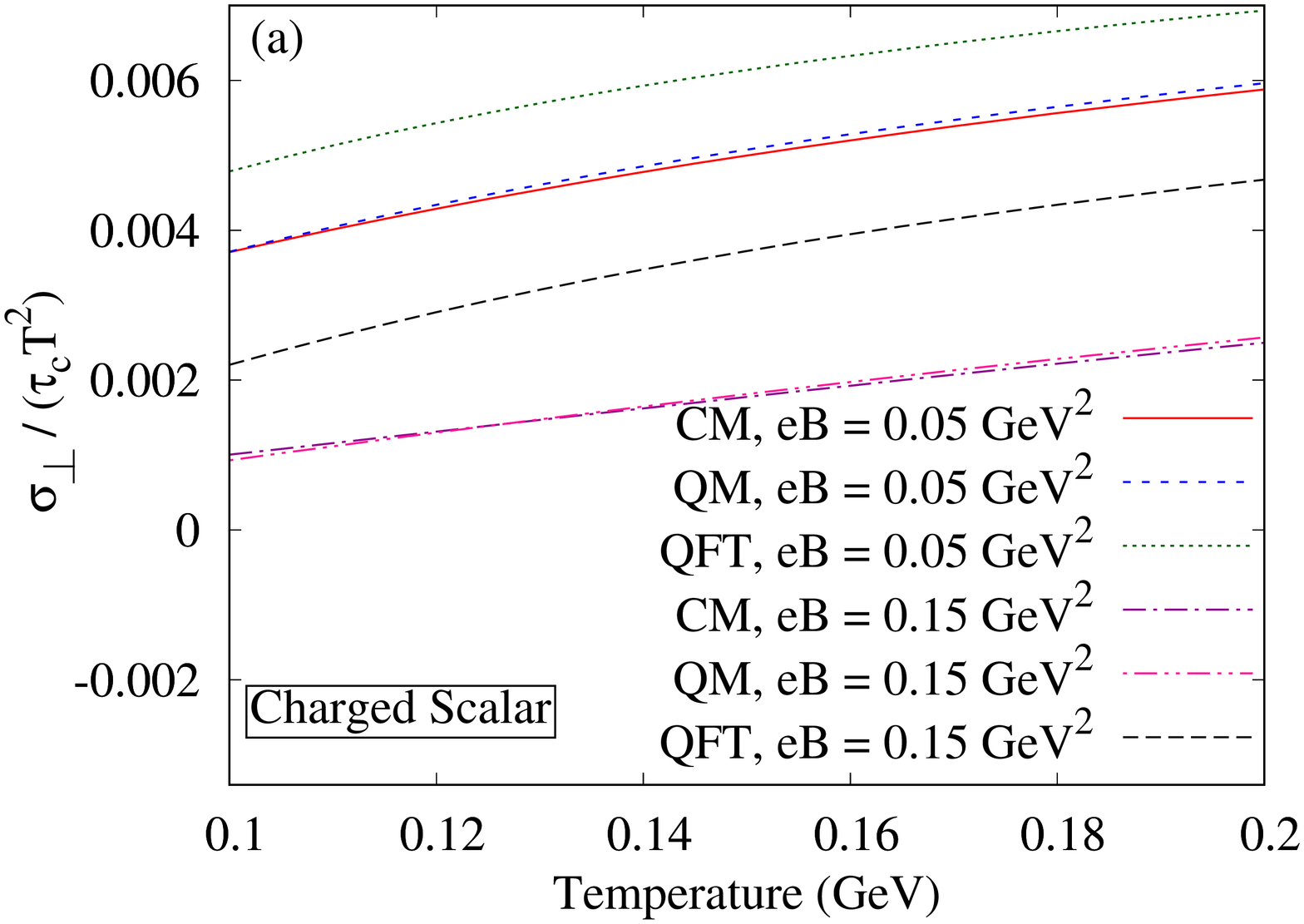}  \includegraphics[angle=-90,scale=0.30]{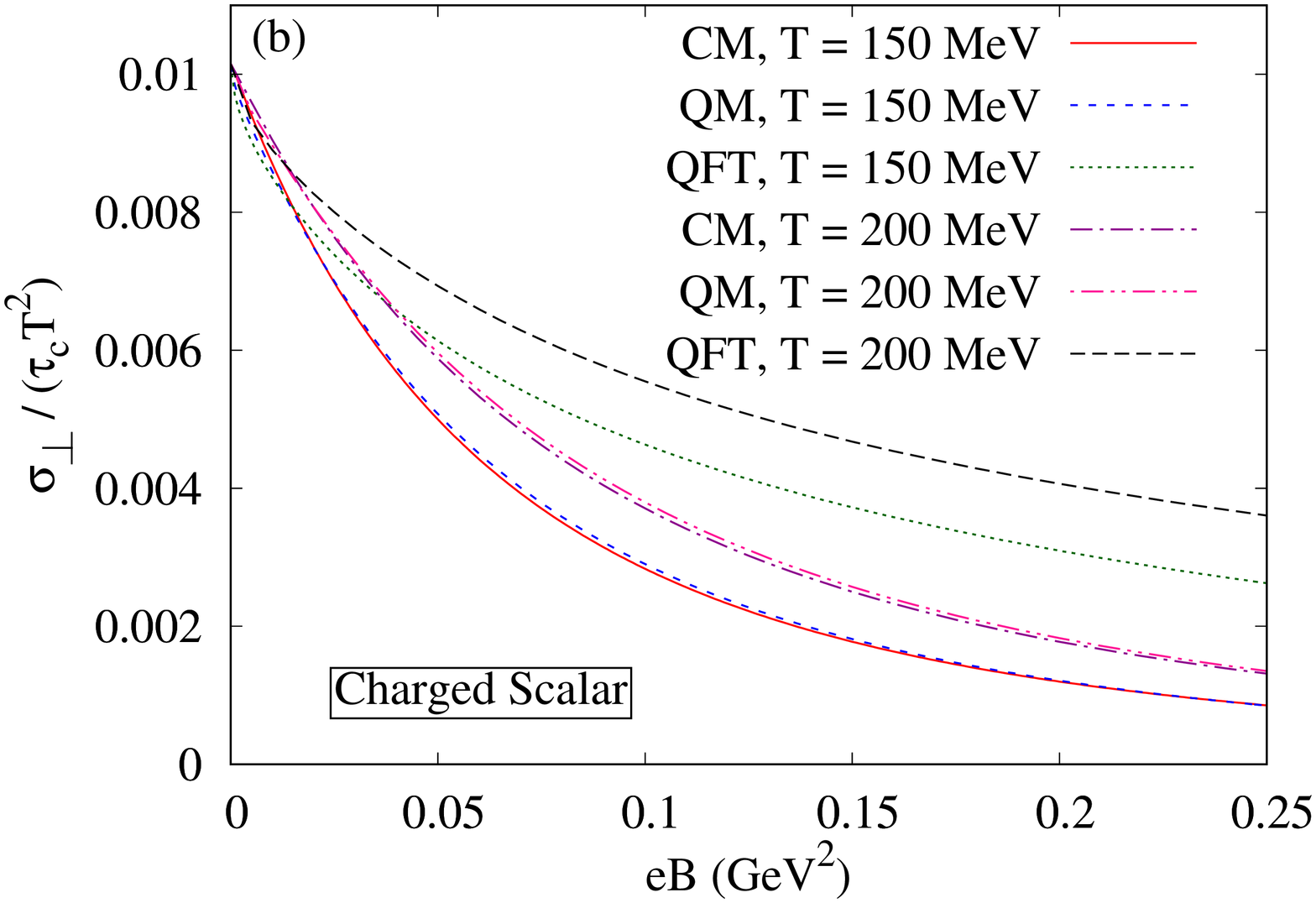}
	\end{center}
	\caption{(Color Online) Normalized values of perpendicular conductivity for CM, QM and QFT expressions of medium with charged scalar particles are plotted as a function of (a) temperature and (b) magnetic field, where $\Gamma^{-1}=1$ fm is considered.}
		\label{fig1}
\end{figure}

\begin{figure}  [h]
	\begin{center}
		\includegraphics[angle=-90,scale=0.30]{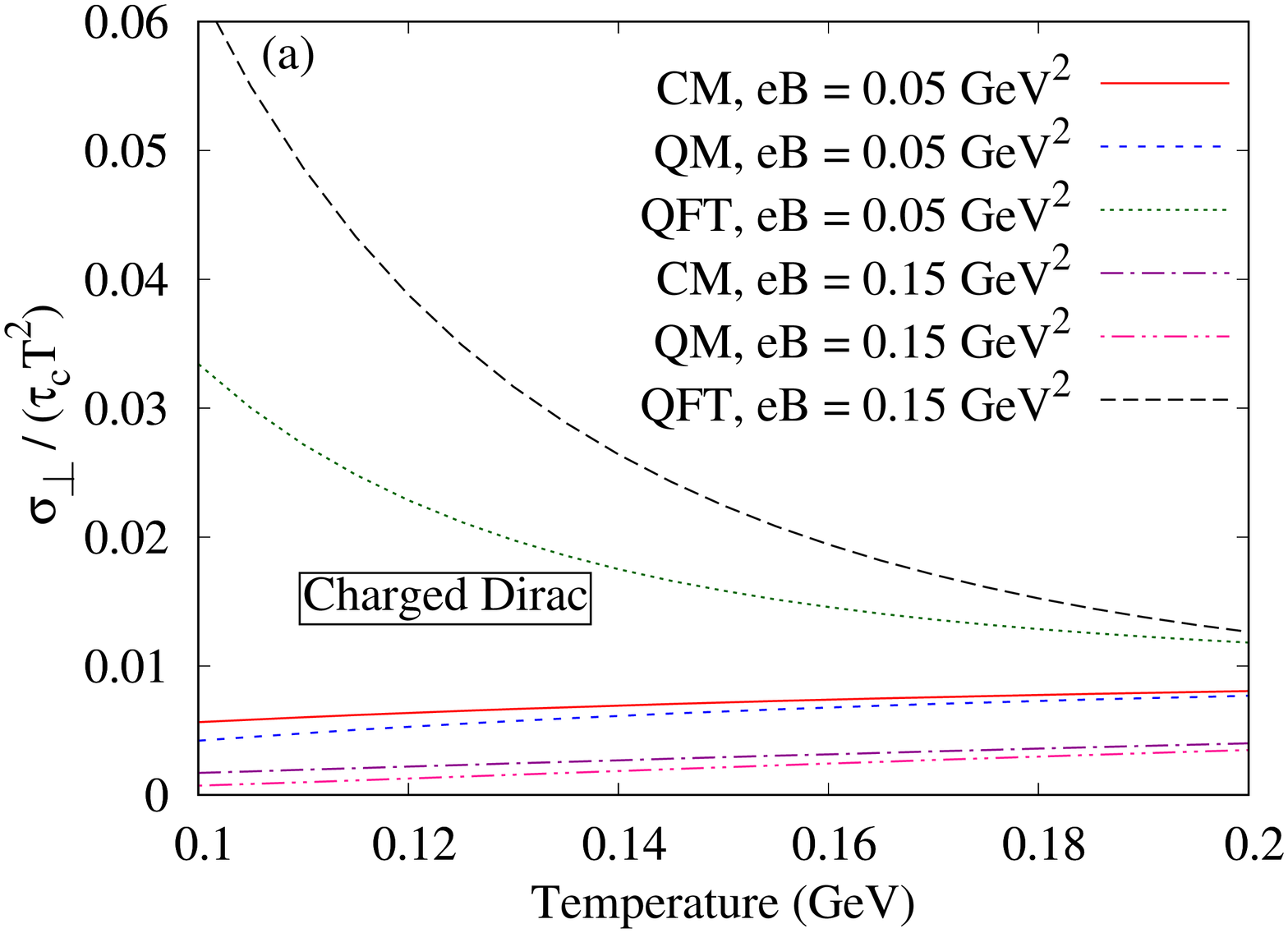}  \includegraphics[angle=-90,scale=0.30]{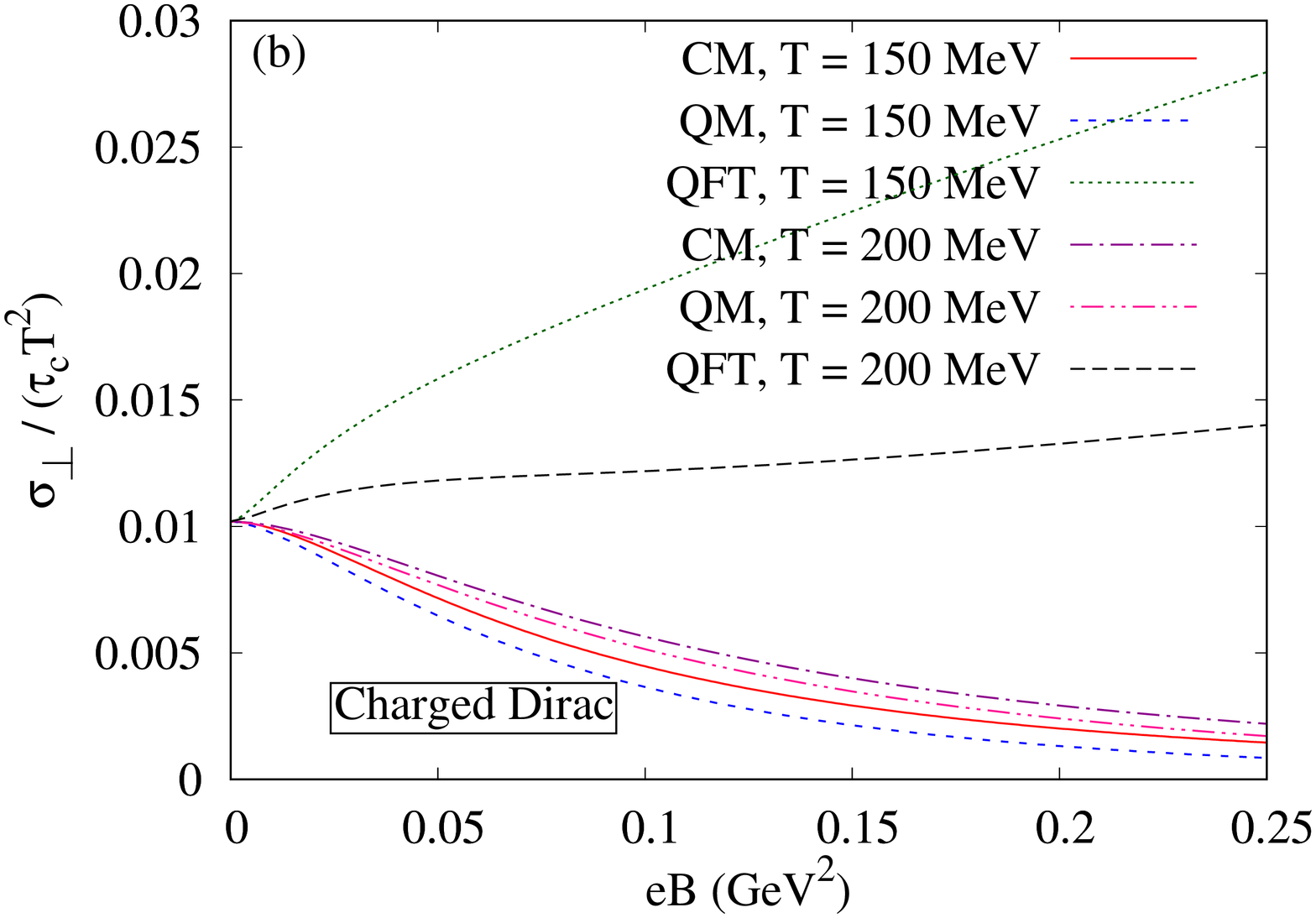}
	\end{center}
	\caption{(Color Online) Same as Fig.~(\ref{fig1}) for medium with charged Dirac particles.}
		\label{fig2}
\end{figure}

Before discussing the various numerical graphs and results, we first specify here the methods for the numerical calculations. The final expression of $\sigma^\parallel$ and $\sigma^\perp$ contain sum over Landau levels and integral over momentum. The numerical integration was performed using the CQUAD routine of GSL library~\cite{GSL} in C++. The sum is performed using standard C++ library. For all the results presented here, we have taken upto 10,000 Landau levels. Figs.~(\ref{fig1}) and (\ref{fig2}) show the variation of $\sigma^{\perp}$ of scalar Bosons and Dirac Fermions with $B$ and $T$. We have considered dimensionless quantity $\sigma^{(\perp)}/(\tau_c T^2)$ as in absence of magnetic field, we get almost temperature independent value of $\sigma/(\tau_c T^2)$. So, being a $T$, $B$ independent dimensionless quantity for field-free picture, its variation with $T$ and $B$ for finite magnetic field picture probably provide us a good visualization. That is why we have chosen this dimensionless quantity.  

Let us first see CM curve (red solid line, violet dash-dotted line) of $\sigma^{\perp}$ as a function of $T$ and $B$, shown in Figs.~\ref{fig1}(a) and (b) respectively. If we notice the RTA expression of Eq.~(\ref{RTA_B0}), then one can identify the extra factor/fraction $1/\Big(1 + \frac{\tau_c^2}{\tau_B^2}\Big)$, for which the normalized conductivity component $\sigma^{(\perp)}/(\tau_c T^2)$ gets an additional $T$, $B$ dependence. It can be checked by imposing $B\rightarrow 0$ limit, where the factor/fraction will be merged to unity. Hence, $T$, $B$ dependence of $1/\Big(1 + \frac{\tau_c^2}{\tau_B^2}\Big)\approx 1/\Big(1 + \frac{e^2B^2\tau_c^2}{9T^2}\Big)$ (considering average $\tau_B= \frac{\langle\omega\rangle }{eB} \simeq \frac{3T}{eB}$ since $\langle\omega\rangle \simeq 3T$ in accordance with the law of equipartition of energy for massless particles) is basically reflected in the CM curves, drawn in the Figs.~\ref{fig1}(a) and (b). 
Hence, by analyzing that factor, one can get the mathematical reason for increasing and decreasing trends of $\sigma^{(\perp)}/(\tau_c T^2)$ in $T$ and $B$-axes respectively. The physical reason may be qualitatively understood by the following way. Increasing temperature means increasing randomness or disorders while magnetic field is responsible for alignment of system, meaning less randomness or more orders. This opposite effect of $T$ and $B$ on many body system may be considered as the hidden physics, for which an opposite trends of $\sigma^{(\perp)}/(\tau_c T^2)$ in $T$ and $B$-axes are noticed. It is the RTA based Boltzmann transport equation in kinetic theory framework, which carries this opposite roles of $T$ and $B$ through its detailed mathematical anatomy. Now, when we go to the QM curves - blue short-dash and pink double dash-dotted lines, then additional Landau level summation effect will come into the picture, whose effect is appeared to be very mild because CM and QM curves are approximately merged with each other. In zoom-in view, one can find a mild noticeable difference between them. 

For getting actual quantum effect, we should trust more on QFT curves (green dotted and black long dash lines) because QM curves contain mixture of quantum (Landau quantization) and classical (cyclotron motion) information. If we compare the RTA expressions of $B=0$ and $B\neq 0$, given in Eqs.~(\ref{RTA_B0}) and (\ref{cond_RTAB}) respectively then we get the main (anisotropic) factor $1/\Big(1 + \frac{\tau_c^2}{\tau_B^2}\Big)$ in $\sigma^{(\perp)}_\text{Scalar/Dirac}$, for which entire changes in CM expressions and major changes in QM expressions are occurred. In QFT expression, given in Eq.~(\ref{condgeneral}), one can identify similar kind of factor but in different form: $\dfrac{\Gamma}{\Gamma^2 + (\omega_{kl}-\omega_{kn})^2}$.  Using the standard relation between the relaxation time ($\tau_c$) and thermal width $\Gamma$ as $\tau_c  = \frac{1}{\Gamma}$ and imposing an equivalence between Landau level transition ($\omega_{kl} - \omega_{kn}$) and cyclotron frequency $1/\tau_B = \frac{eB}{\omega}$ as $\omega_{kl} - \omega_{kn}\equiv\frac{1}{\tau_B}$, we may build a connection  $\dfrac{\Gamma}{\Gamma^2 + (\omega_{kl}-\omega_{kn})^2} \approx \dfrac{\tau_c}{1 + \frac{\tau_c^2}{\tau_B^2}}$. Although this is not exactly that straight forward mapping as it is written. Its more delicate mapping is hidden in different Kronecker deltas, which will be discussed in next paragraph.   
This interesting transformation from classical to quantum picture has also been explored in Ref.~\cite{Ghosh:2020wqx} for viscosity of relativistic fluid in presence of magnetic field. If we collectively consider present work on conductivity and Ref.~\cite{Ghosh:2020wqx} on viscosity, then we notice a beautiful (common) physics in transportation, which reveals a transformation from the cyclotron frequency in CM picture to the transition between Landau levels in QFT picture. 

Now, we find that QFT curves are quite larger than QM and CM curves. The reason might be understood if we analyze all terms of Kronecker deltas in $\mathcal{A}_{ln}^{(2)}$, given in Eq.~(\ref{eq.A2}). They are basically restricting the values of $l$ and $n$, which might be compared with the selection rules appeared in hydrogen atom problem. The $\delta_{ln}$, $\delta_{l,n+1}$ and $\delta_{l,n-1}$ will fix $l=n$, $l=n+1$ and $l=n-1$ options, which means that two virtual scalar particles of current-current correlator can have three possible modes of transportation with differences of Landau levels $\Delta_{ln}=0$, $+1$, $-1$ respectively. Interestingly, first term with same Landau levels option will give us the QM expression $\sigma^\text{QM}_{\perp}$, given in Eq.~(\ref{Lsig_QM}) but excluding the anisotropic fraction/factor $1/\Big[1+\Big(\frac{\tau_c}{\tau_B}\Big)^2\Big]$. Now, second and third terms have factors $\dfrac{\Gamma}{\Gamma^2 + (\omega_{k(n+1)}-\omega_{kn})^2}$ and $\dfrac{\Gamma}{\Gamma^2 + (\omega_{k(n-1)}-\omega_{kn})^2}$, which might be equivalent to the anisotropic factor $\tau_c/\Big[1+(\frac{\tau_c}{\tau_B})^2\Big]$ of CM expressions if we trust on the  equivalence : $(\omega_{k(n-1)}-\omega_{kn})^2\equiv(\omega_{k(n+1)}-\omega_{kn})^2\equiv1/\tau_B^2$. Since this equivalence works dimensionally but not exactly, so in one direction we may be happy with this equivalence realization but in other direction, we should deal with exact mathematical anatomy by accepting the transformations from CM to QM to QFT.

Next in Figs.~\ref{fig2}(a) and (b), $\sigma^{(\perp)}/(\tau_cT^2)$ for a medium with Dirac spin-$1/2$ particles are plotted as a function os $T$ and $B$ respectively. Similar to scalar particle medium, CM (red solid and violet dash-dotted lines) and QM (blue short-dash and pink dash-double-dotted lines) curves are mainly controlled by the anisotropic factor $\tau_c/\Big[1+(\frac{\tau_c}{\tau_B})^2\Big]$, for which a mild enhancement with $T$ and a noticeable reduction with $B$ are observed. Now when we go for QFT curve, a sudden enhancement is observed. Similar to scalar case, To realize this enhancement of QFT version of conduction, we have to focus on Kronecker deltas in $\mathcal{D}_{ln}^{(0)}$, given in Eq.~(\ref{eq.Z1}).
The $\mathcal{D}_{ln}^{(0)}$ contains $\delta_{l,n+1}$ and $\delta_{l,n-1}$, which fix $l=n+1$ and $l=n-1$ options and they mean that two virtual Dirac particles of current-current correlator can have two possible modes of transportation with differences of Landau levels $\Delta_{ln}=+1$ and $-1$ respectively. Unlike to scalar case, Dirac case is not carrying any $\delta_{ln}$, which can be equal to its QM expression of Eq.~(\ref{Lsig_QM}) but excluding the anisotropic fraction/factor $1/\Big[1+\Big(\frac{\tau_c}{\tau_B}\Big)^2\Big]$. 
Another interesting point is that with respect to the conductivity at $B=0$, QFT version of perpendicular conductivity for scalar and Dirac cases are going lower and higher direction respectively.    %
\begin{figure}[h]
	\begin{center}
		\includegraphics[angle=-90,scale=0.30]{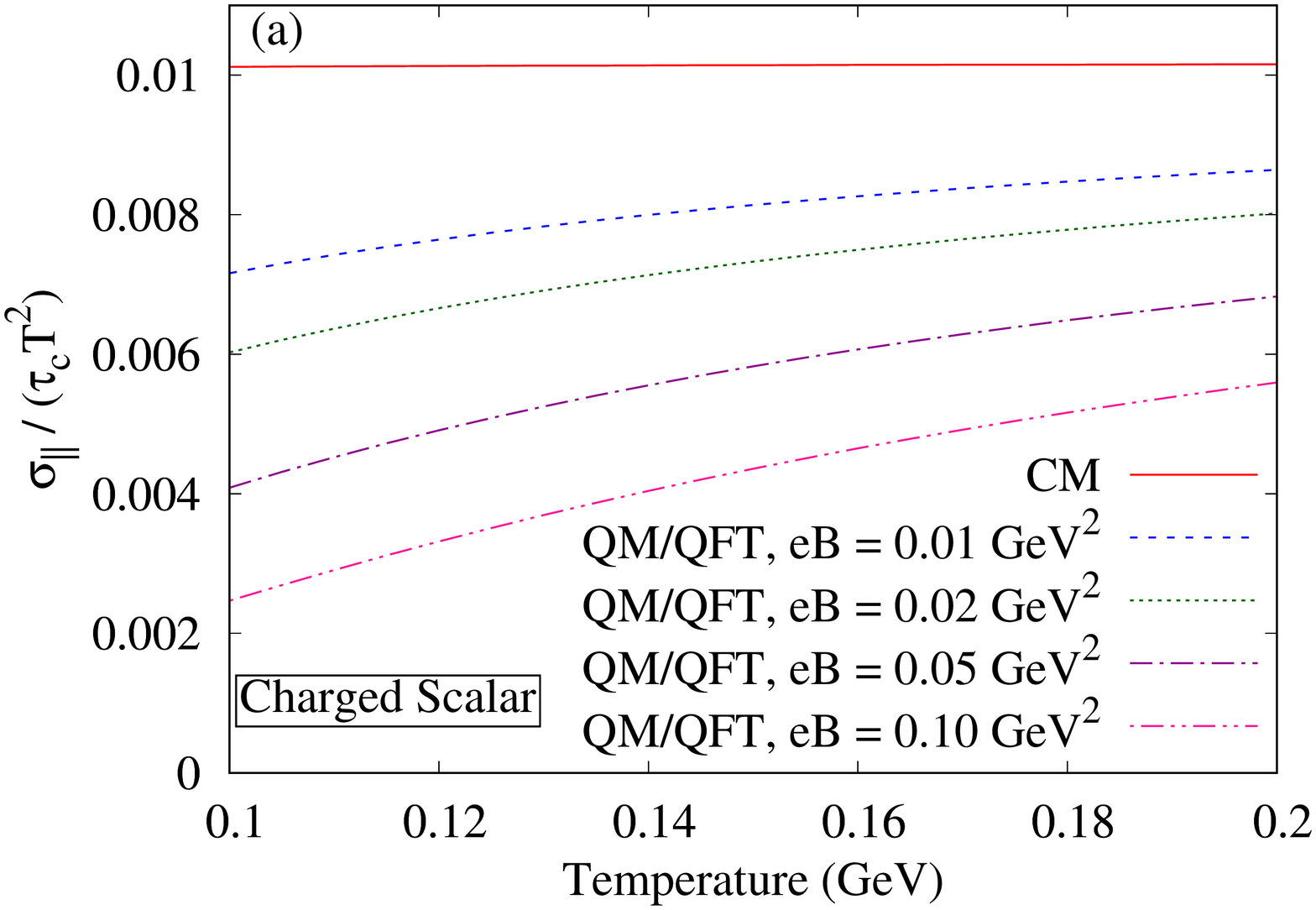}  \includegraphics[angle=-90,scale=0.30]{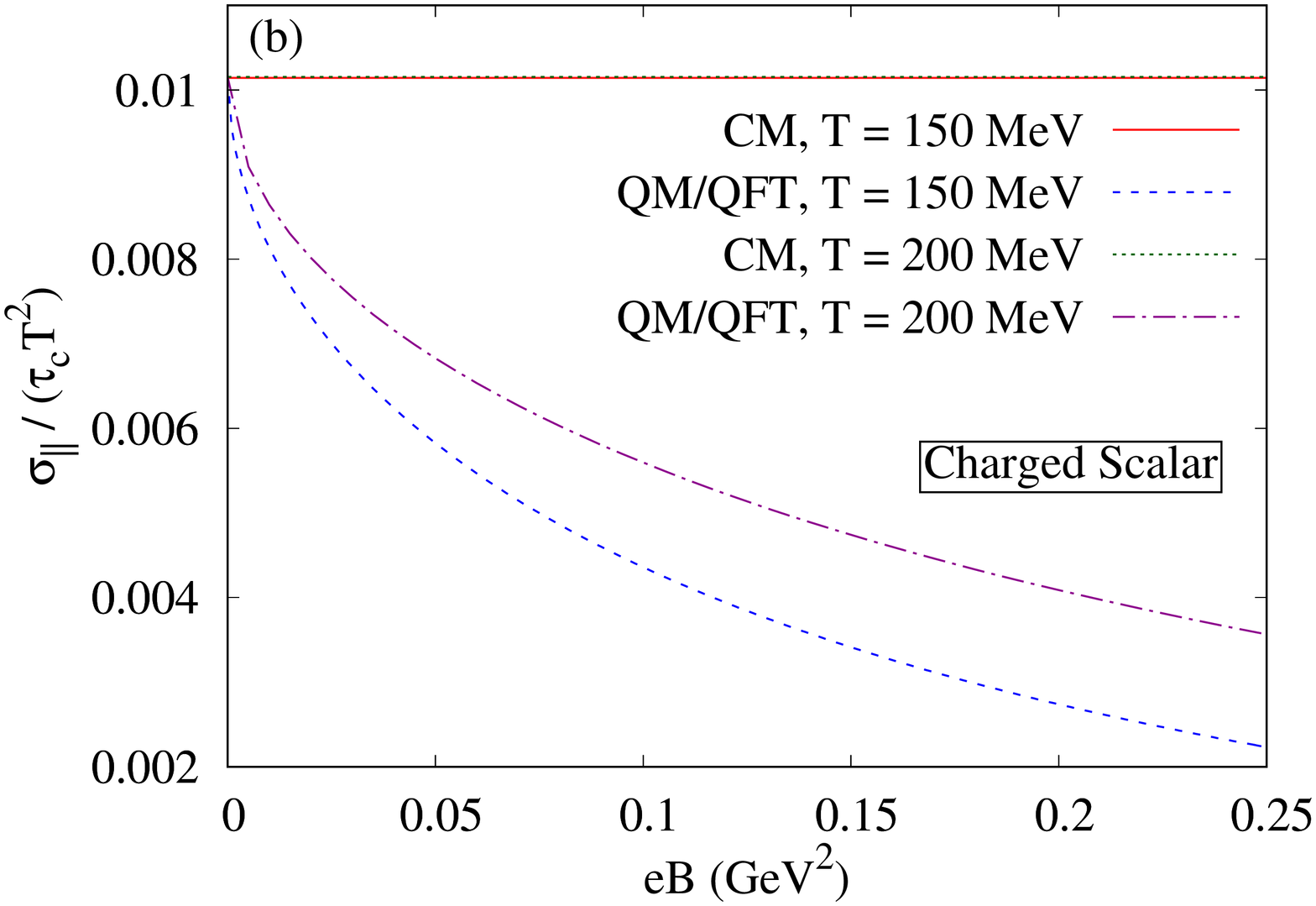}
		\includegraphics[angle=-90,scale=0.30]{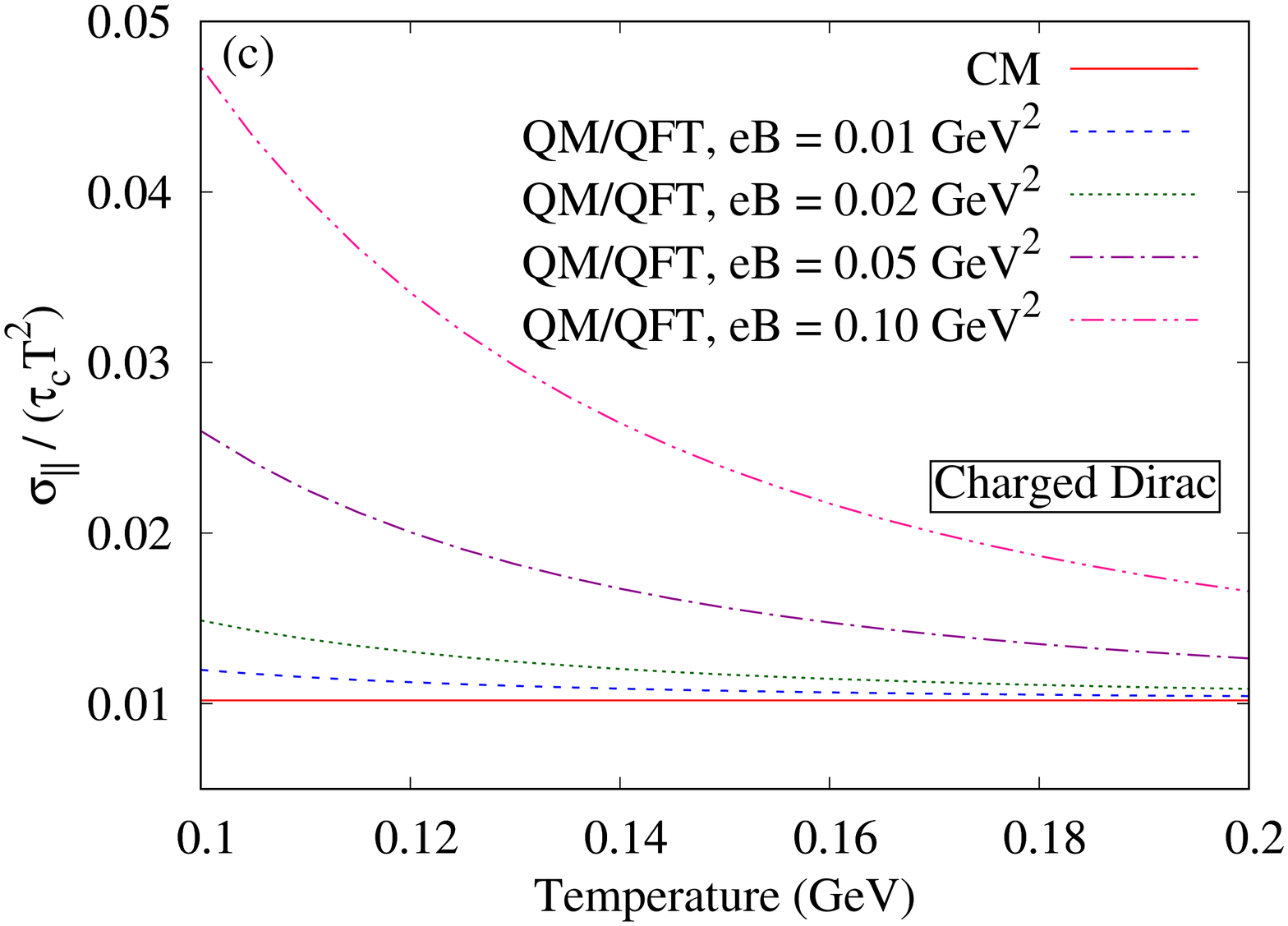}  \includegraphics[angle=-90,scale=0.30]{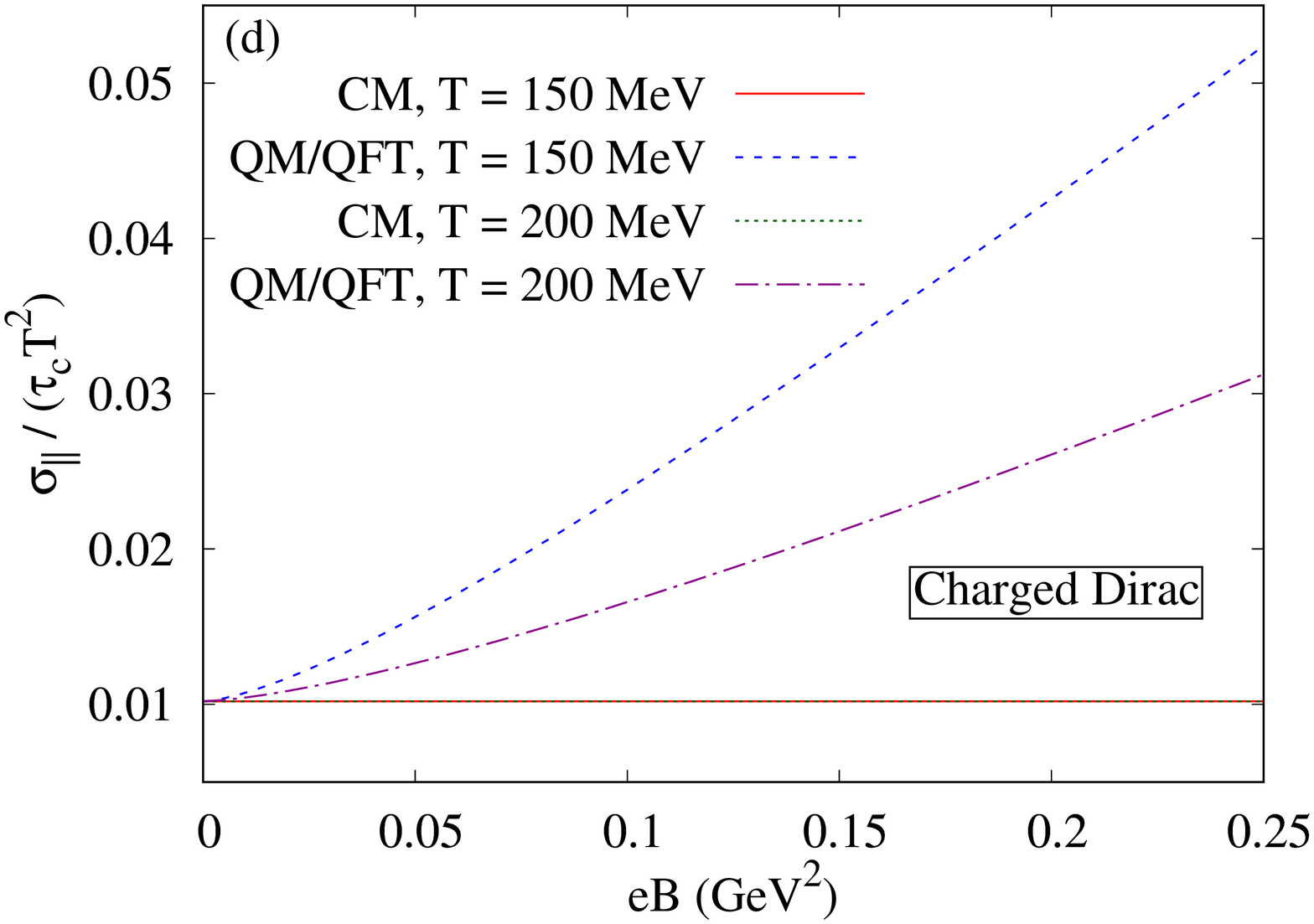}
	\end{center}
	\caption{(Color Online) (a) $T$ and (b) $eB$-dependence of (normalized) parallel conductivity for CM, QM and QFT expressions of medium with scalar particles. (c) $T$ and (d) $eB$-dependence of (normalized) parallel conductivity for CM, QM and QFT expressions of medium with Dirac particles. In all the plots, $\Gamma^{-1}=1$ fm is considered.}
	\label{fig3}
\end{figure}

In Figs.~\ref{fig3}(a)-(d), we have shown the variation of the quantity $\sigma_{||}/\tau_cT^2$ as a function of $T$ and $eB$ for scalar and Dirac cases. The CM curves of normalized parallel conductivity $\sigma_{||}/\tau_cT^2$ is independent of $T$ and $B$, which will be very interesting reference point for quantum version.
We know that the Lorentz force remain handicap along the direction of $B$, that is why parallel conductivity become $B$ independent. Red horizontal lines in upper panels (for scalar case) and lower panel (for Dirac case) reveal this fact. Now it is because of Landau quantization in quantum picture, this component get some $B$-dependent structure as well as the deviation from $T^2$-dependence. The interesting result of Fig.~\ref{fig3} is that QM and QFT plots are the same i.e. $\sigma^\pll_\text{QFT} = \sigma^\pll_\text{QM}$ which has been explicitly shown in Appendix~\ref{app.d}. With respect to $B=0$ result, parallel conduction of scalar and Dirac medium become respectively smaller and larger at finite $B$. Origin of this deviation from $B=0$ value of parallel conductivity is completely quantum in nature. By zooming the B-axis, one can find a lower values of B, beyond which this deviation is started.

It is to be noted that, in this work, we have considered only finite temperature whereas the chemical potential $\mu=0$. We find that, at $\mu=0$, the de Haas-Van Alphen (dHVA) oscillation~\cite{LANDAU1980158,Fukushima:2007fc,Noronha:2007wg,Ebert:1998gx} are missing in the Fermionic spectral function. However, we have also checked that, at high $\mu$ and low $T$, the spectral function possess dHVA oscillation. It will be interesting to analyze the $\mu$ dependence of the conductivities at low temperature and high magnetic field, relevant for magnetars which can be an immediate extension of the current work.
\section{SUMMARY AND CONCLUSIONS}
In summary, we have explored field theoretical calculation of conductivity for relativistic fluid in presence of magnetic field. With the help of rich anatomy of propagators of scalar and Fermion at finite temperature and magnetic field, the detail calculation of conductivity components for scalar and Fermionic medium have been done. Owing to the famous Kubo relation, electrical conductivity can be connected with zero momentum limit of current-current correlator. In absence of magnetic field, the expression of Kubo approach is exactly equal to  the expression, based on relaxation time approximation (RTA) method. If we analyze their final expressions, then we can identify two parts - phases space factor with Fermi-Dirac (FD) or Bose-Einstein (BE) distribution function and dissipation/interaction part. The latter part in RTA is taking care by relaxation time $\tau_c$, which measure the time scale of the medium for approaching from equilibrium to non-equilibrium. In Kubo approach, this part is taken care by spectral representation $\FB{\dfrac{\Gamma}{(\om_{q,k}-\om_k)^2+\Gamma^2}}$ of two virtual scalar or Fermion particles (quantum fluctuation at finite temperature), whose zero (external) momentum limit ($q\rightarrow 0$) give us $\frac{1}{\Gamma}=\tau_c$, which reveal the equivalence role by thermal width or scattering probability $\Gamma$ of medium constituents in Kubo framework as played by $\tau_c$ in RTA. Now, in presence of magnetic field, the conductivity tensor become anisotropic, which means that perpendicular and parallel components of conductivity with respect to applied external magnetic field will be different (which were same in absence of magnetic field). An interesting point at finite $B$ picture is that the RTA and Kubo expressions of conductivity components are not remain same. To understand this discrepancy in terms of physical interpretation, we have marked RTA as CM expression and Kubo as QFT expression as Landau quantization is missing in former. By incorporating Landau quantization in RTA, we build an intermediate expression, by naming QM expression. Interestingly QM and QFT expressions are coming same for parallel component but it is not true for perpendicular component. Probably our imposing Landau quantization in RTA for perpendicular component still remain in semi-classical in nature, as it is till carrying classical quantity $\tau_B$. QFT expressions provide us a full quantum description, where classical information of $\tau_B$ is transformed to (inverse of) energy difference between two Landau levels. Similar to selection rule in hydrogen atom problem, here we get three possible Landau level differences: 
$\Delta_{n,l}=0$, $+1$, $-1$, imposed through Kronecker deltas in final expressions. They are coming due to orthogonal properties of Laggure polinomials, representing quantum mechanical probabilistic existences of two virtual particles. Our calculations are basically expose the different modes of quantum fluctuations at finite temperature and magnetic field, where those modes are fixed by orthogonal properties of their quantum probabilistic (Laggure polynomials) functions. Via quantum field theory, our outcome of parallel and perpendicular components of scalar and Dirac medium can be written in bullet points as follows:
\begin{enumerate}[(i)]
    \item Parallel component of conductivity for scalar medium comes from quantum fluctuation at finite $T$ and $B$ with same Landau level ($l=n$) internal lines (i.e. $\Delta_{n,l}=0$) and effective relaxation time scale $\tau_c=1/\Gamma$.
    \item Parallel component of conductivity for Dirac medium comes from quantum fluctuation at finite $T$ and $B$ with same Landau level internal lines (i.e. $\Delta_{n,l}=0$) and effective relaxation time scale $\tau_c=1/\Gamma$. Here same Landau level restriction comes through two Kronecker deltas $\delta_{l,n}$ and $\delta_{l-1,n-1}$, whose combination help beautifully to build spin degeneracy factor $g_n=2-\delta_{n,0}$ for $n$-th Landau level.
    \item Perpendicular component of conductivity for scalar medium comes from quantum fluctuation at finite $T$ and $B$ with two internal lines, having Landau level differences $\Delta_{n,l}=0$, $+1$ and $-1$ and effective relaxation time scale $\tau_c=1/\Gamma$, $\dfrac{\Gamma}{\Gamma^2 + (\omega_{k(n+1)}-\omega_{kn})^2}$ and $\dfrac{\Gamma}{\Gamma^2 + (\omega_{k(n-1)}-\omega_{kn})^2}$ respectively $\Big( {\rm where}~ \om_{kn}=\sqrt{k_z^2+(2n+1)eB}\Big)$.
    \item Perpendicular component of conductivity for Dirac medium comes from quantum fluctuation at finite $T$ and $B$ with two internal lines, having Landau level differences $\Delta_{n,l}=+1$ and $-1$ and effective relaxation time scale $\tau_c=\dfrac{\Gamma}{\Gamma^2 + (\omega_{k(n+1)}-\omega_{kn})^2}$ and $\dfrac{\Gamma}{\Gamma^2 + (\omega_{k(n-1)}-\omega_{kn})^2}$ respectively $\Big( {\rm where}~ \om_{kn}=\sqrt{k_z^2+2neB}\Big)$.
\end{enumerate}

At the end, our investigation might have lot of scope for future studies to get more mature understanding and exploration. Our immediate future plan is to use this field theoretical expressions of parallel and perpendicular component conductivity to RHIC or LHC matter, which can face a strong magnetic field. Since we have taken a general massless relativistic Fermionic system, we could not directly compare our results of Sec.~\ref{sec.numerics} with Lattice QCD. Now, considering a 2-flavour quark matter, if we multiply the spectral function by $N_c \sum_{f} Q_f^2 = \frac{5}{3}$, then we get at $T \simeq 250$ MeV (at high temperature, quarks may be treated as massless), the electrical conductivity $\sigma \simeq 5.3~\tau$ MeV/fm. Considering, $\tau \sim $ few fm, (which is realistic for hot QCD medium), the magnitude of the conductivity comes out to be in agreement with the Lattice QCD estimate of Ref.~\cite{Ding:2010ga}, though the relaxation time $\tau$ (inverse of thermal width $\Gamma$) in our calculation is a parameter put by hand. It will be interesting to calculate the electrical conductivity in QGP medium using the Kubo formalism by considering the effect of realistic `strong' interaction in $\Gamma$ (or $\tau$) in near future.

\section*{ACKNOWLEDGEMENTS} 
SS thanks to Amaresh Jaiswal and Guruprasad Kadam for providing a visiting fellowship to NISER, Orissa, funded from DST-INSPIRE faculty with Grant No. DST/INSPIRE/04/2017/000038. Snigdha Ghosh is funded by the Department of Higher Education, Government of West Bengal. We would also like to thank Prof. Faqir Khanna and Guruprasad Kadam for valuable discussions and useful comments on the manuscript. 
 

\appendix
\section{CALCULATION OF THE VECTOR CURRENT CORRELATORS} \label{app.1}
In this appendix, we explicitly evaluate the two-point vector current correlation function $\ensembleaverage{J^{\mu}(x)J^{\nu}(y)}_{11}$ for scalar as well as Dirac theory. Let us first consider the Bosonic case. From Eq.~(\ref{Current_Scalar}) we get,
\begin{eqnarray}
\ensembleaverage{\mathcal{T}_cJ^{\mu}_\text{Scalar}(x)J^{\nu}_\text{Scalar}(y)}_{11} && = 
-e^2\ensembleaverage{\mathcal{T}_c \Big(\phi^{\dagger}(x) \partial^{\mu}\phi(x)-\partial^{\mu}\phi^{\dagger}(x) \phi(x)\Big)
 \Big(\phi^{\dagger}(y) \partial^{\nu}\phi(y)-\partial^{\nu}\phi^{\dagger}(y)\phi(y)\Big)}_{11}
\end{eqnarray}
which on applying the Wick's theorem yields the following expression
\begin{eqnarray}
\ensembleaverage{\mathcal{T}_cJ^{\mu}_\text{Scalar}(x)J^{\nu}_\text{Scalar}(y)}_{11} &=& -e^2 \TB{
\wick
{
	\ensembleaverage{\mcTc \c2\phi^\dagger(x)\del^\mu \c1\phi(x)  \c1\phi^\dagger(y)\del^\nu\c2 \phi(y) }_{11} 
- 	\ensembleaverage{\mcTc \c2\phi^\dagger(x)\del^\mu \c1\phi(x)  \del^\nu \c1  \phi^\dagger(y) \c2 \phi(y) }_{11} } \right. \nn \\ && \left. 
\wick
{
	- \ensembleaverage{\mcTc \del^\mu \c2\phi^\dagger(x) \c1\phi(x)  \c1\phi^\dagger(y)\del^\nu\c2 \phi(y) }_{11} 
	+ 	\ensembleaverage{\mcTc \del^\mu \c2\phi^\dagger(x) \c1\phi(x) \del^\nu \c1  \phi^\dagger(y) \c2 \phi(y) }_{11} }
}
\label{A2}
\end{eqnarray}
Simplifying the above equation, we arrive at
\begin{eqnarray}
\ensembleaverage{\mathcal{T}_cJ^{\mu}_\text{Scalar}(x)J^{\nu}_\text{Scalar}(y)}_{11} &=& -e^2\Big[\partial^{\mu}_xD_{11}(x,y)\partial_y^{\nu}D_{11}(y,x) - \partial_x^{\mu}\partial_y^{\nu}D_{11}(x,y) D_{11}(y,x) \nn\\
&& - D_{11}(x,y)\partial^{\mu}_x\partial^{\nu}_yD_{11}(y,x) + \partial^{\nu}_yD_{11}(x,y) \partial^{\mu}_xD_{11}(y,x)\Big]
\label{A3}
\end{eqnarray}
where, $\del^\mu_x \equiv \frac{\del}{\del x_\mu}$, $\del^\mu_y \equiv \frac{\del}{\del y_\mu}$ etc. and $D_{11}(x,y) = \wick{\ensembleaverage{\mcTc \c\phi(x) \c\phi^\dagger(y) }_{11}}$ denotes the 11-component of the coordinate space thermal scalar propagator in RTF. Owing to the translation invariance of the propagator $D_{11}(x,y)=D_{11}(x-y)$, it can be Fourier transformed to momentum space as
\begin{eqnarray}
D_{11}(x,y) = D_{11}(x-y) = \int\!\!\!\frac{d^4k}{(2\pi)^4}e^{-ik\cdot(x-y)}(-iD_{11}(k;m))
\label{A4}
\end{eqnarray}
in which, $D_{11}(k;m)$ denotes the corresponding 11-component of the free thermal scalar propagator in momentum space whose explicit form reads~\cite{Mallik:2016anp,Bellac:2011kqa} 
\begin{eqnarray}
D_{11}(k;m) = \frac{-1}{k^2-m^2+i\epsilon} + \xi_+(k.u)2\pi i\delta(k^2-m^2).
\label{A5}
\end{eqnarray}
In the above equation, $u^{\mu}$ is the four-velocity of the thermal bath which becomes $u^\mu_\text{LRF}\equiv(1,\vec{0})$ in the Local Rest Frame (LRF) of the bath, $\xi_+(x) = \Theta(x)f_+(x) + \Theta(-x)f_+(-x) $ and $f(x) =\TB{e^{x/T}-1}^{-1}$
denotes the Bose-Einstein thermal distribution function. 
Making use of Eq.~(\ref{A4}) into Eq. (\ref{A3}), we get, 
\begin{eqnarray}
\ensembleaverage{\mathcal{T}_cJ^{\mu}_\text{Scalar}(x)J^{\nu}_\text{Scalar}(y)}_{11} = -\int\!\!\!\int\!\!\!\frac{d^4p}{(2\pi)^4}\frac{d^4k}{(2\pi)^4}e^{-i(x-y)\cdot(p-k)}
D_{11}(p;m)D_{11}(k;m)\mathcal{N}^{\mu\nu}_\text{Scalar}(k,p)
\label{A6}
\end{eqnarray}
where,
\begin{eqnarray}
\mathcal{N}^{\mu\nu}_\text{Scalar}(k,p) = e^2\FB{p^{\mu}k^{\nu}  + k^{\mu}p^{\nu} + p^{\mu}p^{\nu} + k^{\mu}k^{\nu}}. 
\label{A7}
\end{eqnarray}
For the calculation of the electrical conductivity tensor, we need the expression of  $\mathcal{N}^{\mu\nu}_\text{Scalar}(k,k)$ which is obtained from the above equation as  
\begin{eqnarray}
\mathcal{N}^{\mu\nu}_\text{Scalar}(k,k) = 4e^2k^{\mu}k^{\nu}.
\label{A8}
\end{eqnarray}


Following a similar procedure, we can now calculate the corresponding two-point vector current correlation function $\ensembleaverage{\mathcal{T}_c J^{\mu}_\text{Dirac}(x)J^{\nu}_\text{Dirac}(y)}_{11}$ for Dirac case. From Eq.~(\ref{Current_Scalar}) we get,
\begin{eqnarray}
\ensembleaverage{\mathcal{T}_c J^{\mu}_\text{Dirac}(x)J^{\nu}_\text{Dirac}(y)}_{11}  =
e^2\ensembleaverage{\mathcal{T}_c\psibar(x)\gamma^{\mu}\psi(x)\psibar(y)\gamma^{\nu}\psi(y)}_{11}\nn \\
&&\end{eqnarray}
which on applying the Wick's theorem gives 
\begin{eqnarray}
\ensembleaverage{\mathcal{T}_c J^{\mu}_\text{Dirac}(x)J^{\nu}_\text{Dirac}(y)}_{11} &=& 
e^2 \wick[offset=1.2em]{ \ensembleaverage{\mcTc \c2\psibar(x)\gamma^\mu \c1 \psi(x) \c1 \psibar(y)\gamma^\nu \c2 \psi(y)}_{11}} \nn \\
&=& -e^2\Tr\SB{\gamma^{\mu}S_{11}(x,y)\gamma^{\nu}S_{11}(y,x)}.
\label{A9}
\end{eqnarray}
In Eq.~\eqref{A9}, $S_{11}(x,y) = \wick[offset=1.2em]{\ensembleaverage{\mcTc \c\psi(x) \c\psibar(y) }_{11}}$ denotes the 11-component of the coordinate space thermal Dirac propagator in RTF. It is interesting to note that, the above expression remains valid even if the Fermion field $\psi$ is a multiplet. In that case, the traces have to be taken over all the spaces belonging to the multiplet along with the Dirac space. As the thermal Dirac propagator $S_{11}(x,y)=S_{11}(x-y)$ is translationally invariant, it can be Fourier transformed to momentum space as
\begin{eqnarray}
S_{11}(x,y) = S_{11}(x-y) = \int\!\!\!\frac{d^4p}{(2\pi)^4}e^{-ip\cdot(x-y)}\FB{-iS_{11}(p;m)}
\label{A10}
\end{eqnarray}
where, $S_{11}(p;m)$ denotes the corresponding 11-component of the free thermal Dirac propagator in momentum space, explicitly given by the following expression~\cite{Mallik:2016anp,Bellac:2011kqa}
\begin{eqnarray} 
S_{11}(p;m) = \FB{\cancel{p} + m}\tilde{D}_{11}(p;m)
\label{1011}
\end{eqnarray}
with, 
\begin{eqnarray}
\tilde{D}_{11}(p;m) = \TB{\frac{-1}{p^2-m^2+i\epsilon} - \xi_-(p.u)2\pi i\delta(p^2-m^2)}.
\label{1012}
\end{eqnarray} 
In the above equation, $\xi_-(x)=\Theta(x)f_-(x)+\Theta(-x)f_-(-x)$ and 
$f_-(x)=\TB{e^{x/T}+1}^{-1}$ denotes the Fermi-Dirac thermal distribution function.
On substituting Eq.~\eqref{A10} into Eq.~(\ref{A9}), we obtain 
\begin{eqnarray}
\ensembleaverage{\mathcal{T}_c J^{\mu}_\text{Dirac}(x)J^{\nu}_\text{Dirac}(y)}_{11} && = 
e^2\int\!\!\!\int\!\!\!\frac{d^4p}{(2\pi)^4}\frac{d^4k}{(2\pi)^4}e^{-i(x-y)\cdot (p-k)}
\Tr\SB{\gamma^{\mu}S_{11}(p;m)\gamma^{\nu}S_{11}(k;m)}\label{A11.1} \\
&& = - \int\!\!\!\int\!\!\!\frac{d^4p}{(2\pi)^4}\frac{d^4k}{(2\pi)^4}e^{-i(x-y)\cdot(p-k)}\tilde{D}_{11}(p;m)\tilde{D}_{11}(k;m)\mathcal{N}^{\mu\nu}_\text{Dirac}(k,p) 
\label{A11}
\end{eqnarray}
where, 
\begin{eqnarray}
\mathcal{N}^{\mu\nu}_\text{Dirac}(k,p) &=& -e^2\Tr\SB{ \gamma^{\mu}(\cancel{p}+m)\gamma^{\nu}(\cancel{k}+m)} \nn \\ 
&=& -4e^2 \TB{p^\mu k^\nu + k^\mu p^\nu - g^\munu \SB{(k\cdot p)^2-m^2}}.
\label{A12}
\end{eqnarray}
For the calculation of the electrical conductivity tensor, we need the expression of  $\mathcal{N}^{\mu\nu}_\text{Dirac}(k,k)$ which is obtained from the above equation as  
\begin{eqnarray}
\mathcal{N}^{\mu\nu}_\text{Dirac}(k,k) = -4e^2\TB{2k^{\mu}k^{\nu} - g^{\mu\nu}(k^2-m^2)}.
\label{A13}
\end{eqnarray}


\section{CALCULATION OF THE VECTOR CURRENT CORRELATORS IN PRESENCE OF EXTERNAL MAGNETIC FIELD} \label{app.B}
Here, we give the calculation of the two point vector current correlation function  $\ensembleaverage{\mathcal{T}_cJ^{\mu}(x)J^{\nu}(y)}_{11}^{B}$ in the presence of
an external magnetic field. For this, we will proceed along the same lines as the $B=0$ case so that in presence of magnetic field, Eq.~\eqref{A3} for the scalar filed becomes   
\begin{eqnarray}
\ensembleaverage{\mathcal{T}_cJ^{\mu}_\text{Scalar}(x)J^{\nu}_\text{Scalar}(y)}_{11} &=& -e^2\Big[D^{\mu}_xD_{11}(x,y)D_y^{\nu}D_{11}(y,x) - D_x^{\mu}D_y^{*\nu}D_{11}(x,y) D_{11}(y,x) \nn\\
&& - D_{11}(x,y)D^{*\mu}_xD^{\nu}_yD_{11}(y,x) + D^{*\nu}_yD_{11}(x,y) D^{*\mu}_xD_{11}(y,x)\Big]
\label{eq.B1}
\end{eqnarray}
where, 
$D^\mu_x \equiv \del^\mu_x + ieA_\text{ext}^\mu(x)$, 
$D^{*\mu}_x \equiv \del^\mu_x - ieA_\text{ext}^\mu(x)$ etc.  
and $D^B_{11}(x,y) = \wick{\ensembleaverage{\mcTc \c\phi(x) \c\phi^\dagger(y) }^B_{11}}$ denotes the 11-component of the coordinate space free thermo-magnetic scalar propagator in RTF. Unlike the $B=0$ case, the thermo-magnetic propagator is not translationally invariant $D^B_{11}(x,y)=\Phi(x,y)D^B_{11}(x-y)$ as it contains the gauge dependent phase factor $\Phi(x,y)$ which is responsible for explicitly breaking the translational invariance. The translationally invariant piece $D^B_{11}(x-y)$ of the propagator can be Fourier transformed to the momentum space as 
\begin{eqnarray}
D_{11}^B(x-y) = \int\!\!\!\frac{d^4k}{(2\pi)^4}e^{-ik\cdot(x-y)}\FB{-iD_{11}^B(k;m)}
\label{eq.B2}
\end{eqnarray}
where $D^B_{11}(k;m)$ is the corresponding 11-component of the momentum space free thermo-magnetic scalar propagator in RTF whose explicit form is~\cite{Ayala:2004dx}
\begin{eqnarray}
D_{11}^B(k;m) = \sum_{l=0}^{\infty}2(-1)^le^{-\alpha_k}L_l(2\alpha_k)D_{11}(k_\parallel,m_l).
\label{B2}
\end{eqnarray}
In the above equation, $l$ is the Landau level index, $\alpha_k = -\frac{k_{\perp}^2}{eB} \geq 0$, $m_l = \sqrt{m^2 + (2l+1)eB}$ is the 
``\textit{Landau level dependent effective mass}'' and $D_{11}$ is given in Eq.~\eqref{A5}. As we have considered the external magnetic field to be in the $\hat{z}$-direction, we  decompose any four vector $k^\mu$ as $k=(\kpll+\kper)$ where $\kpll^\mu=\gpll^\munu k_\nu$ and $\kper^\mu=\gper^\munu k_\nu$ in which the corresponding decomposition of the metric tensor is $g^\munu=(\gpll^\munu+\gper^\munu)$ with  $\gpll^\munu=\text{diag}(1,0,0,-1)$ and $\gper^\munu=\text{diag}(0,-1,-1,0)$. It is to be noted that, in our convention, $\kper^\mu$ is a space-like vector with $\kper^2=-(k_x^2+k_y^2) <0$ in contrary to the convention used in Ref.~\cite{Ayala:2004dx}.

Although, the thermo-magnetic propagator contains the phase factor $\Phi(x,y)$ (which breaks the translational invariance), the two-point correlation function $\ensembleaverage{\mathcal{T}_cJ^{\mu}_\text{Scalar}(x)J^{\nu}_\text{Scalar}(y)}_{11}^B$ comes out to be translationally invariant. To see this, we take the following explicit form of the phase factor~\cite{Ayala:2004dx} 
\begin{eqnarray}
\Phi(x,y) = \exp\TB{ie\int_{x}^{y}dx'_{\mu}A^{\mu}_\text{ext}(x')}
\label{B3}
\end{eqnarray}
and differentiate it separately with respect to $x$ and $y$ using Leibniz rule to obtain
\begin{eqnarray}
\del^\mu_x \Phi(x,y) &=& \Phi(x,y) \SB{-ie A_\text{ext}^\mu(x)}, \\
\del^\mu_y \Phi(x,y) &=& \Phi(x,y) \SB{ie A_\text{ext}^\mu(x)}
\end{eqnarray}
which in turn yields the following result
\begin{eqnarray}
D^{\mu}_x\Phi(x,y) = D^{*\mu}_y\Phi(x,y) = 0.
\label{B6}
\end{eqnarray}
From Eq.~(\ref{B6}), we obtain 
\begin{eqnarray}
D^\mu_x D^B_{11}(x,y)&=& D^\mu_x \TB{\Phi(x,y)D^B_{11}(x-y)} = \Phi(x,y)\del^\mu_xD^B_{11}(x-y), 
\label{B7}\\
D^{*\mu}_y D^B_{11}(x,y)&=& D^{*\mu}_y \TB{\Phi(x,y)D^B_{11}(x-y)} = \Phi(x,y)\del^\mu_yD^B_{11}(x-y).
\label{B8}
\end{eqnarray}
We now use Eqs.~\eqref{B7} and \eqref{B8} in order to simplify Eq.~\eqref{eq.B1} and get,
\begin{eqnarray}
\ensembleaverage{\mathcal{T}_cJ^{\mu}_\text{Scalar}(x)J^{\nu}_\text{Scalar}(y)}_{11} &=& -e^2 \Phi(x,y)\Phi(y,x) \Big[\partial^{\mu}_xD_{11}(x-y)\partial_y^{\nu}D_{11}(y-x) - \partial_x^{\mu}\partial_y^{\nu}D_{11}(x-y) D_{11}(y-x) \nn\\
&& - D_{11}(x-y)\partial^{\mu}_x\partial^{\nu}_yD_{11}(y-x) + \partial^{\nu}_yD_{11}(x-y) \partial^{\mu}_xD_{11}(y-x)\Big].
\label{B11}
\end{eqnarray}
It is easy to see that, the phase factor in Eq.~\eqref{B3} satisfies  $\Phi(x,y)\Phi(y,x)=1$, so that the correlator $\ensembleaverage{\mathcal{T}_cJ^{\mu}_\text{Scalar}(x)J^{\nu}_\text{Scalar}(y)}_{11}$ of the above equation is translationally invariant and gauge independent. The cancellation of the phase factor for the one-loop diagrams containing equally charged particles is well known~\cite{Ayala:2004dx,Ghosh:2019fet,Ghosh:2018xhh}. Comparing Eq.~\eqref{B11} with Eq.~\eqref{A3}, we notice that the expression of the vector current correlator at $B\ne0$ is identical to the same at $B=0$, except the thermal propagator has been replaced by the translationally invariant piece of the thermo-magnetic propagator.

Substituting Eq.~\eqref{eq.B2} into Eq.~\eqref{B11} and simplifying, we arrive at 
\begin{eqnarray}
\ensembleaverage{\mathcal{T}_cJ^{\mu}_\text{Scalar}(x)J^{\nu}_\text{Scalar}(y)}_{11} = 
-\sum_{l=0}^{\infty}\sum_{n=0}^{\infty}\int\!\!\!\int\!\!\!\frac{d^4p}{(2\pi)^4}\frac{d^4k}{(2\pi)^4}e^{-i(x-y)\cdot (p-k)}D_{11}(k_\parallel;m_l)D_{11}(p_\parallel,m_n)
\mathcal{N}^{\mu\nu}_{ln;\text{Scalar}}(k,p)
\label{B12}
\end{eqnarray}
where 
\begin{eqnarray}
\mathcal{N}_{ln;\text{Scalar}}^{\mu\nu}(k,p) = 4(-1)^{l+n}e^{-\alpha_k-\alpha_p}L_l(2\alpha_k)L_n(2\alpha_p)\mathcal{N}^{\mu\nu}_\text{Scalar}(k,p)
\label{eq.N.Scalar}
\end{eqnarray}
with $\mathcal{N}^{\mu\nu}_\text{Scalar}(k,p)$ given in Eq.~\eqref{A7}.

In the calculation of the electrical conductivity tensor, we need the expression of  $\mathcal{N}_{ln;\text{Scalar}}^{\mu\nu}(k,k)$ which is obtained from the above equation as  
\begin{eqnarray}
\mathcal{N}_{ln;\text{Scalar}}^{\mu\nu}(k,k) = 16e^2\mathcal{A}_{ln}(\kper^2)k^\mu k^\nu.
\label{B13}
\end{eqnarray}
in which
\begin{eqnarray}
\mathcal{A}_{ln}(\kper^2) &=& (-1)^{l+n}e^{-2\alpha_k} L_l(2\alpha_k) L_n(2\alpha_k).
\label{eq.Anl}
\end{eqnarray}


The calculation of the vector current correlator $\ensembleaverage{\mcTc J_\text{Dirac}^\mu(x)J_\text{Dirac}^\nu(y)}^B_{11}$ 
for the Dirac case at $B\ne0$ can be done in a similar way to the scalar case. Here, Eq.~\eqref{A11.1} at $B\ne0$ becomes
\begin{eqnarray}
\ensembleaverage{\mcTc J_\text{Dirac}^\mu(x)J_\text{Dirac}^\nu(y)}^B_{11}  &=&
e^2\int\!\!\!\int\!\!\!\frac{d^4p}{(2\pi)^4}\frac{d^4k}{(2\pi)^4}e^{-i(x-y)\cdot (p-k)}
\Tr\SB{\gamma^{\mu}S^B_{11}(p;m)\gamma^{\nu}S^B_{11}(k;m)} 
\label{B15}
\end{eqnarray}
where, 
$S^B_{11}(p)$ is the 11-component of the momentum space free thermo-magnetic Dirac propagator in RTF, explicitly given by~\cite{Ayala:2003pv,Schwinger:1951nm}
\begin{eqnarray}
S_{11}^B(p;m) = \sum_{l=0}^{\infty}(-1)^le^{-\alpha_p}\mathscr{D}_l(p)\tilde{D}_{11}(p_\parallel,m_l)
\label{B151}
\end{eqnarray}
in which $m_l=\sqrt{m^2+(2l+1-2s)eB} = \sqrt{m^2+2leB}$ (for Dirac case, $s=1/2$), $\Dtil_{11}$ is given in Eq.~\eqref{1012} and $\scrD_l(p)$ is 
\begin{eqnarray}
\scrD_l(p) = \FB{\cancel{p}_\parallel+m}\TB{\FB{\mathds{1}+i\gamma^1\gamma^2}L_l(2\alpha_p) 
	- \FB{\mathds{1}-i\gamma^1\gamma^2}L_{l-1}(2\alpha_p)} - 4\cancel{p}_\perp L^1_{l-1}(2\alpha_p)
\label{eq.Dl}
\end{eqnarray}
with the convention $L_{-1}(z) = L_{-1}^1(z) = 0$.

Substituting Eq.~\eqref{B151} into Eq.~\eqref{B15} and simplifying, we get,
\begin{eqnarray}
\ensembleaverage{\mcTc J_\text{Dirac}^\mu(x)J_\text{Dirac}^\nu(y)}^B_{11} &=& 
-\sum_{l=0}^{\infty} \sum_{n=0}^{\infty}\int\!\!\!\int\!\!\!\frac{d^4p}{(2\pi)^4}\frac{d^4k}{(2\pi)^4} e^{-i(x-y)\cdot(p-k)}
 \Dtil_{11}(\ppll;m_n)\Dtil_{11}(\kpll;m_l) 
\mcN_{ln;\text{Dirac}}^\munu(k,p)
\label{eq.B18}
\end{eqnarray}
where, 
\begin{eqnarray}
\mathcal{N}^{\mu\nu}_{ln;\text{Dirac}}(k,p) = -e^2(-1)^{l+n}e^{-\alpha_k-\alpha_p}
\Tr \SB{\gamma^\mu \scrD_n(p) \gamma^\nu \scrD_l(k)}
\label{B16}
\end{eqnarray}

In the calculation of the electrical conductivity tensor, we need the expression of  $\mathcal{N}_{ln;\text{Dirac}}^{\mu\nu}(k,k)$ which is obtained from the above equation as  
\begin{eqnarray}
\mathcal{N}_{ln;\text{Dirac}}^{\mu\nu}(k,k) = -e^2 (-1)^{l+n}e^{-2\alpha_k}
\mcT^\munu_{ln}(k)
\label{eq.N.Dirac}
\end{eqnarray}
where, 
\begin{eqnarray}
\mcT^\munu_{ln}(k) &=& (-1)^{l+n}e^{-2\alpha_k} \Tr \TB{ \gamma^\mu \scrD_n(k)\gamma^\nu \scrD_l(k) } \nn \\
&=&  8\Big[ 8(2\kper^\mu\kper^\nu-\kper^2g^\munu)\mathcal{B}_{ln}(\kper^2)
+ \big\{2\kpll^\mu\kpll^\nu-\gpll^\munu(\kpll^2-m^2)\big\}\mathcal{C}_{ln}(\kper^2) \nn \\ &&
+ (\kpll^2-m^2)\gper^\munu \mathcal{D}_{ln}(\kper^2) + 2(\kpll^\mu\kper^\nu+\kper^\mu\kpll^\nu)\mathcal{E}_{ln}(\kper^2)
\Big], \label{eq.T.munu}
\end{eqnarray}
in which,
\begin{eqnarray}
\mathcal{B}_{ln}(\kper^2) &=& (-1)^{l+n}e^{-2\alpha_k} L^1_{l-1}(2\alpha_k) L^1_{n-1}(2\alpha_k) ~,
\label{eq.Bnl}\\
\mathcal{C}_{ln}(\kper^2) &=& (-1)^{l+n}e^{-2\alpha_k} 
\SB{L_{l-1}(2\alpha_k) L_{n-1}(2\alpha_k) + L_{l}(2\alpha_k) L_{n}(2\alpha_k)} \label{eq.Cnl}~, \\
\mathcal{D}_{ln}(\kper^2) &=& (-1)^{l+n}e^{-2\alpha_k} 
\SB{L_{l}(2\alpha_k) L_{n-1}(2\alpha_k) + L_{l-1}(2\alpha_k) L_{n}(2\alpha_k)} \label{eq.Dnl}~, \\
\mathcal{E}_{ln}(\kper^2) &=& (-1)^{l+n}e^{-2\alpha_k} 
\SB{L_{l-1}(2\alpha_k) L^1_{n-1}(2\alpha_k) - L_{l}(2\alpha_k) L^1_{n-1}(2\alpha_k) \right. \nn \\ && \left.
	+~ L^1_{l-1}(2\alpha_k) L_{n-1}(2\alpha_k)- L^1_{l-1}(2\alpha_k) L_{n}(2\alpha_k)}. \label{eq.Enl}
\end{eqnarray}
Substitution of Eq.~\eqref{eq.T.munu} into Eq.~\eqref{eq.N.Dirac} yields after some simplifications,
\begin{eqnarray}
\mathcal{N}^{\mu\nu}_{ln;\text{Dirac}}(k,k) &=& -8e^2\Big[8\FB{2k_{\perp}^{\mu}k_{\perp}^{\nu} - k_{\perp}^2g^{\mu\nu}}\mathcal{B}_{ln}(\kper^2)  + \SB{ 2k_\parallel^{\mu}k_\parallel^{\nu}-\gpll^{\mu\nu}(k_\parallel^2-m^2)}\mathcal{C}_{ln}(\kper^2)\nn \\
&& + g_{\perp}^{\mu\nu}\FB{k_\parallel^2-m^2}\mathcal{D}_{ln}(\kper^2)+ 2\FB{k_\parallel^{\nu}k_{\perp}^{\mu} + k_\parallel^{\mu}k_{\perp}^{\nu}}\mathcal{E}_{ln}(\kper^2)\Big].
\label{B18}
\end{eqnarray}


\section{ANALYTIC EXPRESSIONS OF $\mathcal{A}_{ln}^{(j)}$, $\mathcal{B}_{ln}^{(j)}$, $\mathcal{C}_{ln}^{(j)}$, $\mathcal{D}_{ln}^{(j)}$ AND $\mathcal{E}_{ln}^{(j)}$}
\label{app.c}
Using the orthogonality of the Laguerre polynomials, one can first derive the following integral identities:
\begin{eqnarray}
&&\int\!\! \frac{d^2\kper}{(2\pi)^2}e^{-2\alpha_k} L^1_{l-1}(2\alpha_k)L^1_{n-1}(2\alpha_k)\kper^2
= - \frac{(eB)^2}{16\pi}n\delta_{l-1}^{n-1}, \label{eq.Aln.j.0}\\
&&\int\!\! \frac{d^2\kper}{(2\pi)^2}e^{-2\alpha_k} L_{l}(2\alpha_k)L_{n}(2\alpha_k)\kper^2
= -\frac{(eB)^2}{16\pi}\Big\{(2n+1)\delta_{l}^{n}-(n+1)\delta_{l}^{n+1} 
- n\delta_{l}^{n-1}\Big\}, \\
&&\int\!\! \frac{d^2\kper}{(2\pi)^2}e^{-2\alpha_k} L_{l}(2\alpha_k)L_{n}(2\alpha_k)
= \frac{eB}{8\pi}\delta_{l}^{n}. \label{eq.iden.f}
\end{eqnarray}

Using the above equations, we perform the $d^2\kper$ integrals of Eqs.~\eqref{eq.Aln.j.0}-\eqref{eq.iden.f} and get
\begin{eqnarray}
\mathcal{A}_{ln}^{(0)} &=& \frac{eB}{8\pi}\delta_{l}^{n}, \label{eq.Z0}\\
\mathcal{A}_{ln}^{(2)} &=& -\frac{(eB)^2}{16\pi}\Big\{(2n+1)\delta_{l}^{n}+(n+1)\delta_{l}^{n+1} 
+ n\delta_{l}^{n-1}\Big\}, \label{eq.A2}\\
\mathcal{B}_{ln}^{(2)} &=& - \frac{(eB)^2}{16\pi}n\delta_{l-1}^{n-1}, \label{eq.B22} \\
\mathcal{C}_{ln}^{(0)} &=& \frac{eB}{8\pi}\FB{\delta_{l}^{n}+\delta_{l-1}^{n-1}}, \label{eq.C0} \\
\mathcal{D}_{ln}^{(0)} &=& -\frac{eB}{8\pi}\FB{\delta_{l}^{n-1} + \delta_{l-1}^{n}}.
\label{eq.Z1}
\end{eqnarray}
We also note that, the Kronecker delta function having a negative index is zero (\textit{i.e.} $\delta_{-1}^{-1}=0$). This follows from the convention of the Laguerre polynomials $L_{-1}(z)=L_{-1}^1(z) =0$ used in Eq.~\eqref{eq.Dl}.

\section{EQUITY OF $\sigma^\pll_\text{QFT}$ AND $\sigma^\pll_\text{QM}$}
\label{app.d}
Substituting Eqs.~\eqref{eq.Z0}, \eqref{eq.B22} and \eqref{eq.C0} into Eqs.~\eqref{N11} and \eqref{N3}, we get 
\begin{eqnarray}
\tilde{\mathcal{N}}_{ln;\text{Scalar}}^{\pll}(k_z) &=& 4e^2\FB{\frac{eB}{2\pi}}k_z^2\delta_l^n ,  \label{eq.t1} \\
\tilde{\mathcal{N}}_{ln;\text{Dirac}}^{\pll}(k_z) &=& -4e^2\FB{\frac{eB}{2\pi}}\SB{-2leB\delta_{l-1}^{n-1} + (k_z^2+2leB)\fb{\delta_l^n + \delta_{l-1}^{n-1}}}.
\label{eq.t2}
\end{eqnarray}
Using the fact that $\delta_{-1}^1 = 0$, we have $l\fb{\delta_{l}^n + \delta_{l-1}^{n-1}} = 2l \delta_{l-1}^{n-1}$ and 
$\fb{\delta_l^n + \delta_{l-1}^{n-1}} = \fb{2-\delta_l^0}\delta_l^n$, so that Eq.~\eqref{eq.t2} can be further simplified to
\begin{eqnarray}
\tilde{\mathcal{N}}_{ln;\text{Dirac}}^{\pll}(k_z) = -4e^2\FB{\frac{eB}{2\pi}}k_z^2 (2-\delta_0^l) \delta_l^n .
\label{eq.t3}
\end{eqnarray}
Eqs.~\eqref{eq.t1} and \eqref{eq.t3} can be generally expressed as
\begin{eqnarray}
\tilde{\mathcal{N}}_{ln}^{\pll}(k_z) &=& a 2e^2 g_l \FB{\frac{eB}{2\pi}}k_z^2\delta_l^n,  \label{eq.t4}
\end{eqnarray}
where,
\begin{eqnarray}
g_l = \begin{cases}
2 ~~\text{for Scalar}, \\
2(2-\delta_0^l) ~~\text{for Dirac},
\end{cases}
\end{eqnarray}
is the Landau level dependent degeneracy.

Substituting Eq.~\eqref{eq.t4} into Eq.~(\ref{condgeneral}), and evaluating one of the double sums using the Kronecker delta function, we get
\begin{eqnarray}
	\sigma^{\pll}_\text{QFT} &=&\frac{e^2}{T}\FB{\frac{eB}{2\pi}}\sum_{l=0}^{\infty}g_l\int_{-\infty}^{+\infty}\frac{dk_z}{2\pi}\frac{k_z^2}{(\omega_{kl})^2} 
	\frac{1}{\Gamma} f_a(\omega_{kl})\SB{1 + af_a(\omega_{kl}},
	\label{qft-qm}
	\end{eqnarray}
	which is identical to the expression of $\sigma^{\pll}_\text{QM}$ of Eqs.~\eqref{eq.scalar.qm} and \eqref{Lsig_QM}.

\bibliographystyle{apsrev4-1}
\bibliography{Kubo-Conductivity}

\end{document}